\documentclass[twocolumn]{IEEEtran}
\usepackage{amsfonts}
\usepackage{mathrsfs}
\usepackage{amssymb,amsmath}

\usepackage{algorithm}
\usepackage{algorithmic}
\usepackage{amsmath,amssymb,epsfig,graphics,subfigure}
\usepackage{theorem}
\usepackage{array,color}
\usepackage[compress]{cite}

\theoremheaderfont{\normalfont\bfseries}

\begin{document}

\title{Transceiver Design for Dual-Hop Non-regenerative MIMO-OFDM Relay Systems Under Channel Uncertainties}

\author{Chengwen Xing, Shaodan Ma, Yik-Chung Wu, and Tung-Sang Ng, \textsl{Fellow, IEEE}
}
\maketitle

\begin{abstract}

In this paper, linear transceiver design for dual-hop non-regenerative (amplify-and-forward (AF)) MIMO-OFDM systems under channel estimation
errors is investigated. Second order moments of channel estimation
errors in the two hops are first deduced. Then based on the Bayesian
framework, joint design of linear forwarding matrix at the relay and
equalizer at the destination under channel estimation errors is
proposed to minimize the total mean-square-error (MSE) of the output
signal at the destination. The optimal designs for both correlated
and uncorrelated channel estimation errors are considered. The
relationship with existing algorithms is also disclosed. Moreover, this design is extended to the joint design involving source precoder design. Simulation
results show that the proposed design outperforms the design based
on estimated channel state information only.

\end{abstract}

\section{Introduction}
\label{sect:intro}

In order to enhance the coverage of base stations and quality of
wireless links, dual-hop relaying is being considered to be one of
the essential parts for future communication systems (e.g., LTE,
IMT-Adanced, Winner Project). In dual-hop cooperative communication,
relay nodes receive signal transmitted from a source and then
forward it to the destination \cite{Scaglione06}, \cite{Laneman04}.
Roughly speaking, there are three different relay strategies:
decode-and-forward (DF), compress-and-forward (CF) and
amplify-and-forward (AF). Among them, AF strategy is the most
preferable for practical systems due to its low complexity
\cite{Medina07,Tang07,Chae08,Behbahani08,Khoshnevis08}.

On the other hand, for wideband communication, multiple-input
multiple-output (MIMO) orthogonal-frequency-division-multiplexing
(OFDM) has gained a lot of attention in both industrial and academic
communities, due to its high spectral efficiency, spatial diversity
and multiplexing gains \cite{Tse05,Ma05,Chen08,Verde10}. The combination of
AF and MIMO-OFDM becomes an attractive option for enabling
high-speed wireless multi-media services \cite{Hammerstrom07}.

In the last decade, linear transceiver design for various systems
has been extensively investigated because of its low implementation
complexity and satisfactory performance \cite{Tse05},
\cite{Bolcskei06}. For linear transceiver design, minimum
mean-square-error (MMSE) is one of the most important and frequently
used criteria
\cite{Sampath01,Palomar03,Joham05,Serbetli04,Guan08,Rong09,Luo04}.
For example, for point-to-point MIMO and MIMO-OFDM systems, linear
MMSE transceiver design has been discussed in details in
\cite{Sampath01,Palomar03,Joham05}. Linear MMSE transceiver design
for multiuser MIMO systems has been considered in
\cite{Serbetli04,Luo04}. For single carrier AF MIMO relay systems,
linear MMSE forwarding matrix at the relay and equalizer at the
destination are joint designed in \cite{Guan08}. Furthermore, the
linear MMSE transceiver design for dual hop MIMO-OFDM relay systems
based on prefect channel state information (CSI) is proposed in
\cite{Rong09}.

In all the above works, CSI is assumed to be perfectly known.
Unfortunately, in practical systems, CSI must be estimated and
channel estimation errors are inevitable. When channel estimation
errors exist, in general, two classes of designs can be employed:
min-max and stochastic designs. If the distributions of channel
estimation errors are known to be unbounded, stochastic design is
preferred. Stochastic design includes probability-based design and
Bayesian design. In this paper, we focus on Bayesian design, in
which an averaged mean-square-error (MSE) performance is considered.
Recently, Bayesian linear MMSE transceiver design under channel
uncertainties has been addressed for point-to-point MIMO systems
\cite{Zhang08,Ding09} and point-to-point MIMO-OFDM systems
\cite{Rey05}.

In this paper, we take a step further and consider the linear MMSE
transceiver design for dual-hop AF MIMO-OFDM relay systems without
the direct link. For channel estimation in the two hops, both the
linear minimum mean square error and maximum likelihood estimators
are derived, based on which the second order moments of channel
estimation errors are deduced. Using the Bayesian framework, channel
estimation errors are taken into account in the transceiver design
criterion. Then a general closed-form solution for the optimal relay
forwarding matrix and destination equalizer is proposed. Both the
uncorrelated and correlated channel estimation errors are
considered. The relationship between the proposed algorithm and
several existing designs is revealed. Furthermore, the proposed
closed-form solution is further extended to an iterative algorithm
for joint design of source precoder, relay forwarding matrix and
destination equalizer.  Simulation results demonstrate that the
proposed algorithms provide an obvious advantage in terms of data
mean-square-error (MSE) compared to the algorithm based on estimated
CSI only.

We want to highlight that the solution proposed in this paper can be
directly extended to the problem minimizing the weighted MSE.
Various objective metrics such as capacity maximization
and minimizing maximum MSE can be transformed to a weighted MSE
problem with different weighting matrices \cite{Sampath01}. For
clearness of presentation, we only consider a sum MSE minimization
problem. On the other hand, minimizing the transmit power with a QoS
requirement is a different perspective for transceiver design.
Formulating and solving this problem is out of the scope of this
paper.

This paper is organized as follows. System model is presented in
Section~\ref{model}. Channel estimators and the corresponding
covariance of channel estimation errors are derived in
section~\ref{channelestimation}. The optimization problem for
transceiver design is formulated in
Section~\ref{problemformulation}. In
Section~\ref{closedformsolution}, the general optimal closed-form
solution for the relay forwarding matrix and destination equalizer
design problem is proposed. The proposed closed-form solution is
further extended to an iterative algorithm to include the design of
source precoder in Section~\ref{Source_Precoder}. Simulation results are given in Section~\ref{simulation} and finally, conclusions are drawn in
Section~\ref{conclusions}.

The following notations are used throughout this paper. Boldface
lowercase letters denote vectors, while boldface uppercase letters
denote matrices. The notations ${\bf{Z}}^{\rm{T}}$,
${\bf{Z}}^{\rm{H}}$ and ${\bf{Z}}^*$ denote the transpose, Hermitian
and conjugate of the matrix ${\bf{Z}}$, respectively, and
${\rm{Tr}}({\bf{Z}})$ is the trace of the matrix ${\bf{Z}}$. The
symbol ${\bf{I}}_{M}$ denotes the $M \times M$ identity matrix,
while ${\bf{0}}_{M \times N}$ denotes the $M \times N$ all zero
matrix. The notation ${\bf{Z}}^{\frac{1}{2}}$ is the Hermitian
square root of the positive semi-definite matrix ${\bf{Z}}$, such
that ${\bf{Z}}={\bf{Z}}^{\frac{1}{2}}{\bf{Z}}^{\frac{1}{2}}$ and
${\bf{Z}}^{\frac{1}{2}}$ is a Hermitian matrix. The symbol
${\mathbb{E}}\{.\}$ represents the expectation operation.  The
operation ${\rm{vec}}({\bf{Z}})$ stacks the columns of the matrix
${\bf{Z}}$ into a single vector. The symbol $\otimes$ represents
Kronecker product. The symbol $a^+$ means $\max \{0,a\}$. The
notation ${\rm{diag}}[{\bf{A}},{\bf{B}}]$ denotes the block diagonal
matrix with ${\bf{A}}$ and ${\bf{B}}$ as the diagonal elements.

\section{System Model}
\label{model}

\begin{figure}[!ht]
\centering
\includegraphics[width=.45\textwidth]{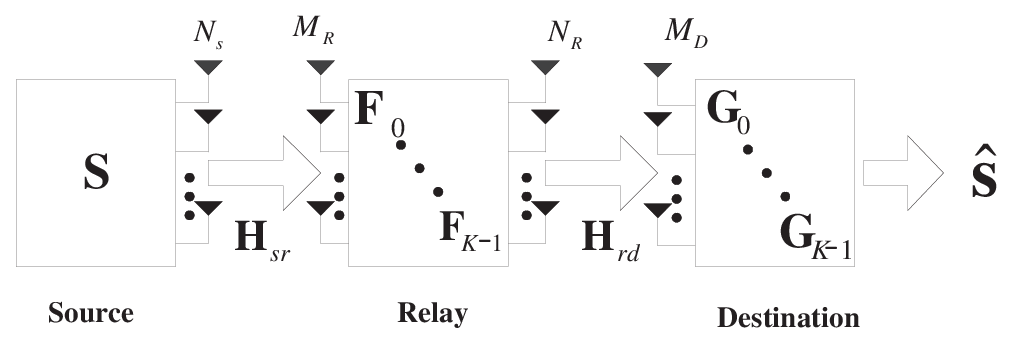}
\caption{Amplify-and-forward MIMO-OFDM relaying diagram.}\label{fig:1}
\end{figure}

In this paper, we consider a dual-hop amplify-and-forward (AF) MIMO-OFDM relaying
cooperative communication system,  which consists of
one source with $N_S$ antennas, one relay with $M_{R}$ receive
antennas and $N_R$ transmit antennas, and one destination with $M_D$
antennas, as shown in Fig.~\ref{fig:1}. At the first hop, the source
transmits data to the relay, and the received signal ${\bf{x}}_k$ at
the relay on the $k^{\rm{th}}$ subcarrier is
\begin{equation}
{\bf{x}}_k= {\bf{H}}_{sr,k}{\bf{s}}_k+{\bf{n}}_{1,k}, \ \ \
k=0,1,\cdots K-1,
\end{equation} where ${\bf{s}}_k$ is the data vector transmitted by
the source with covariance matrix
${\bf{R}}_{{\bf{s}}_k}=\mathbb{E}\{{\bf{s}}_k{\bf{s}}_k^{\rm{H}}\}$
on the $k^{\rm{th}}$ subcarrier, and ${\bf{R}}_{{\bf{s}}_k}$ can be
an arbitrary covariance matrix. The matrix ${\bf{H}}_{sr,k}$ is the
MIMO channel between the source and relay on the $k^{\rm{th}}$
subcarrier. The symbol ${\bf{n}}_{1,k}$ is the additive Gaussian
noise with zero mean and covariance matrix
${\bf{R}}_{n_{1,k}}=\sigma_{n_1}^2{\bf{I}}_{M_R}$ on the
$k^{\rm{th}}$ subcarrier. At the relay, for each subcarrrier, the
received signal ${\bf{x}}_k$ is multiplied by a forwarding matrix
${\bf{F}}_k$, under a power constraint
$\sum_k{\rm{Tr}}({\bf{F}}_k{\bf{R}}_{{\bf{x}}_k}{\bf{F}}_k^{\rm{H}})
\le P_r$ where
${\bf{R}}_{{\bf{x}}_k}=\mathbb{E}\{{\bf{x}}_k{\bf{x}}_k^{\rm{H}}\}$
and $P_r$ is the maximum transmit power. Then the resulting signal
is transmitted to the destination. The received data ${\bf{y}}_k$ at
the destination on the $k^{\rm{th}}$ subcarrier is
\begin{equation}
\label{signal_sep} {\bf{y}}_k= {{\bf{H}}_{rd,k} {\bf{F}}_k
{\bf{H}}_{sr,k}{\bf{s}}}_k + {{\bf{H}}_{rd,k}
{\bf{F}}_k{\bf{n}}_{1,k} } + {\bf{n}}_{2,k},
\end{equation} where the symbol ${\bf{n}}_{2,k}$ is the additive Gaussian noise vector
on the $k^{\rm{th}}$ subcarrier at the second hop with zero mean and
covariance matrix ${\bf{R}}_{n_{2,k}}=\sigma_{n_2}^2{\bf{I}}_{M_D}$.
In order to guarantee the transmitted data ${\bf{s}}_k$ can be
recovered at the destination, it is assumed that $M_R$, $N_R$, and
$M_D$ are greater than or equal to $N_S$ \cite{Behbahani08}.

The signal ${\bf{x}}$ received at the relay and the signal
${\bf{y}}$ received at the destination in frequency domain can be
compactly written as
\begin{align}
 {\bf{x}}&={\bf{H}}_{sr}{\bf{s}}+{\bf{n_1}}, \label{equ:signal_x} \\
 {\bf{y}}&=
{{\bf{H}}_{rd} {\bf{F}} {\bf{H}}_{sr}{\bf{s}}}  + {{\bf{H}}_{rd}
{\bf{F}}{\bf{n}}_{1} } + {\bf{n}}_{2}, \label{equ:signal}
\end{align} where
\begin{subequations}
\begin{align}
& {\bf{y}}\triangleq
[{\bf{y}}_0^{\rm{T}},\cdots,{\bf{y}}_{K-1}^{\rm{T}}]^{\rm{T}}, \ \ \
{\bf{s}}\triangleq [{\bf{s}}_0^{\rm{T}},\cdots,{\bf{s}}_{K-1}^{\rm{T}}]^{\rm{T}} \\
& {\bf{F}}\triangleq {\rm{diag}}[{\bf{F}}_0,\cdots,{\bf{F}}_{K-1}],
\\ & {\bf{H}}_{sr}\triangleq
{\rm{diag}}[{\bf{H}}_{sr,0},{\bf{H}}_{sr,1},\cdots,{\bf{H}}_{sr,K-1}],
\\
&{\bf{H}}_{rd}\triangleq
{\rm{diag}}[{\bf{H}}_{rd,0},{\bf{H}}_{rd,1},\cdots,{\bf{H}}_{rd,K-1}],
\\
&{\bf{n}}_1\triangleq
[{\bf{n}}_{1,0}^{\rm{T}},{\bf{n}}_{1,1}^{\rm{T}},\cdots,{\bf{n}}_{1,K-1}^{\rm{T}}]^{\rm{T}},
\\ & {\bf{n}}_2\triangleq
[{\bf{n}}_{2,0}^{\rm{T}},{\bf{n}}_{2,1}^{\rm{T}},\cdots,{\bf{n}}_{2,K-1}^{\rm{T}}]^{\rm{T}}.
\end{align}
\end{subequations}

Notice that in general the matrix ${\bf{F}}$ in (\ref{equ:signal})
can be an arbitrary $KN_R\times KM_R$ matrix instead of a block
diagonal matrix. This corresponds to mixing the data from different
subcarriers at the relay, and is referred as subcarrier cooperative
AF MIMO-OFDM systems \cite{Rong09}. It is obvious that when the
number of subcarrier $K$ is large, transceiver design for such
systems needs very high complexity. On other hand, it has been shown
in \cite{Rong09} that the low-complexity subcarrier independent AF
MIMO-OFDM systems (i.e., the system considered in
(\ref{equ:signal_x}) and (\ref{equ:signal})) only have a slight
performance loss in terms of total data mean-square-error (MSE)
compared to the subcarrier cooperative AF MIMO-OFDM systems.
Therefore, in this paper, we focus on the more practical subcarrier
independent AF MIMO-OFDM relay systems.

\section{Channel Estimation Error Modeling}
\label{channelestimation}

In practical systems, channel state information (CSI) is unknown and
must be estimated. Here, we consider estimating the channels based
on training sequence. Furthermore, the two frequency-selective MIMO
channels between the source and relay, and that between the relay
and destination are estimated independently. In this work, the
source-relay channel is estimated at the relay, while the
relay-destination channel is estimated at the destination. Then each
channel estimation problem is a standard point-to-point MIMO-OFDM
channel estimation.

For point-to-point MIMO-OFDM systems, channels can be estimated in
either frequency domain or time domain. The advantage of time domain
over frequency domain channel estimation is that there are much
fewer parameters to be estimated \cite{Larsson03}. Therefore, we
focus on time domain channel estimation. Because the channels in the
two hops are separately estimated in time domain, we will present
the first hop channel estimation as an example and the same
procedure can be applied to the second hop channel estimation.

From the received signal model in frequency domain given by
(\ref{equ:signal_x}), the corresponding time domain signal is
\begin{align}
\label{signal_time} {\bf{r}}=&({\boldsymbol{\mathcal{F}}}^{\rm{H}}
\otimes {\bf{I}}_{M_R})
{\bf{x}}\nonumber \\
=&\underbrace{({\boldsymbol{\mathcal{F}}}^{\rm{H}} \otimes
{\bf{I}}_{M_R}) {\bf{H}}_{sr}({\boldsymbol{\mathcal{F}}} \otimes
{\bf{I}}_{N_S})}_{\triangleq{\boldsymbol{\mathcal{H}}}_{sr}}
\underbrace{({\boldsymbol{\mathcal{F}}}^{\rm{H}} \otimes
{\bf{I}}_{N_S}){\bf{s}}}_{\triangleq{\bf{d}}}\nonumber \\
&  + \underbrace{
({\boldsymbol{\mathcal{F}}}^{\rm{H}} \otimes {\bf{I}}_{M_R})
{\bf{n}}_1}_{ \triangleq{\bf{v}}}
\end{align} where ${\boldsymbol{\mathcal{F}}}$ is the normalized discrete-Fourier-transform (DFT)
matrix with dimension $K \times K$.  Based on the properties of DFT
matrix, it is proved in Appendix~\ref{app_sigmal_model} that
(\ref{signal_time}) can be rewritten as
\begin{align}
\label{signalmodel} {\bf{{
r}}}=\left({\bf{D}}^{\rm{T}}\otimes{\bf{I}}_{M_R}\right)\underbrace{{{\rm{vec}}}([
{{\boldsymbol{\mathcal{H}}}_{sr}^{(0)}} \   \cdots  \
{{\boldsymbol{\mathcal{H}}}_{sr}^{(L_1-1)}}])}_{\triangleq
{\boldsymbol{\xi }}_{sr}}+{\bf{v}},
\end{align} where the matrices
${\boldsymbol{\mathcal{H}}}_{sr}^{(\ell)}$ are defined as
\begin{align}
\label{H_sr_time}
{\boldsymbol{\mathcal{H}}}_{sr}^{(\ell)}=\frac{1}{K}\sum_{k=0}^{K-1}{\bf{H}}_{sr,k}e^{j\frac{2\pi}{K}k
\ell}, \ \ \ \ \ell=0,1,\cdots L_1-1.
\end{align} It is obvious that ${\boldsymbol{\mathcal{H}}}_{sr}^{(\ell)}$ is the $\ell^{\rm{th}}$ tap of the multi-path MIMO channel
between the source and relay in the time domain and $L_1$ is the
length of the multi-path channel. The data matrix ${\bf{D}}$ is a
block circular matrix as
\begin{align}
\label{D_Matrix} {\bf{D}}& \triangleq \left[ {\begin{array}{*{20}c}
   {{\bf{d}}_0} &   {{\bf{d}}_1}   &  \cdots  &  \cdots  &  \cdots  & {{\bf{d}}_{K-1}} \\
   {{\bf{d}}_{K-1}} &  {{\bf{d}}_0} &  \ddots  &  \ddots  &  \vdots  & {{\bf{d}}_{K-2}}  \\
    \vdots  &  \cdots  &  \ddots  &  \ddots  &  \vdots  &  \vdots   \\
   {{\bf{d}}_{K-L_1+1}} &  {{\bf{d}}_{K-L_1+2}}  &  \cdots  & \cdots &  \cdots  & {{\bf{d}}_{K-L_1}}  \\
\end{array}} \right],
\end{align} where the element ${\bf{d}}_i$ is expressed as
\begin{align}
{\bf{d}}_i=\frac{1}{\sqrt{K}}
\sum_{k=0}^{K-1}{\bf{s}}_{k}e^{j\frac{2\pi}{K}ki}, \ \ \
i=0,\cdots, K-1.
\end{align}

Based on the signal model in (\ref{signalmodel}), the linear minimum-mean-square-error (LMMSE) channel estimate is given by \cite{Larsson03}
\begin{align}
\label{estimtaion_MMSE} {\boldsymbol{\hat
{\xi}}}_{sr}=&(\sigma_{n_1}^{-2}({\bf{D}}^{\rm{T}}\otimes{\bf{I}}_{M_R})^{\rm{H}}
({\bf{D}}^{\rm{T}}\otimes{\bf{I}}_{M_R})+{\bf{R}}_{\rm{channel}}^{-1})^{-1}\nonumber \\
& \times \sigma_{n_1}^{-2}({\bf{D}}^{\rm{T}}
\otimes{\bf{I}}_{M_R})^{\rm{H}}{\bf{r}},
\end{align} with the corresponding MSE
\begin{align}
\label{MSE_LMMSE}
&\mathbb{E}\{({{\boldsymbol{\xi}}_{sr}-{\boldsymbol{\hat
\xi}}_{sr}})({\boldsymbol{\xi}}_{sr}-{\boldsymbol{\hat
\xi}}_{sr})^{\rm{H}}\}\nonumber \\&=({\bf{R}}_{\rm{channel}}^{-1}+{{\sigma_{n_1}^{-2}}
({\bf{D}}^{*}{\bf{D}}^{\rm{T}})}\otimes{\bf{I}}_{M_R})^{-1},
\end{align}
where ${\bf{R}}_{\rm{channel}}={\mathbb{E}}\{ {\boldsymbol{\xi}}_{sr}{{\boldsymbol{\xi}}_{sr} }^{\rm{H}}\}$ is the prior information for channel covariance matrix. For uncorrelated channel taps,
${\bf{R}}_{\rm{channel}}={\boldsymbol \Lambda}_{\rm{channel}}\otimes{\bf{I}}_{M_RN_S}$ and ${\boldsymbol \Lambda}_{\rm{channel}}={\rm{diag}}[\sigma_{h_0},\sigma_{h_1},\cdots,\sigma_{h_{L-1}}]$, where $\sigma_{h_{l}}$ is the variance of the $l^{\rm{th}}$ channel tap \cite{Rey05}.

On the other hand, the channel in frequency domain and time domain
has the following relationship\footnote{This relationship holds for both perfect CSI and estimated CSI.}
\begin{align}
\label{f_t_relation} {{\rm{vec}}}([ {{{\bf{H}}}_{sr,0}} \ \cdots \
{{\bf{H}}}_{sr,K-1}])=\sqrt{K}({\boldsymbol{\mathcal{F}}}_{L_1}
\otimes {\bf{I}}_{M_RN_S}){\boldsymbol{\xi}}_{sr},
\end{align}where ${\boldsymbol{\mathcal{F}}}_{L_1}$ is the first $L_1$ columns of ${\boldsymbol{\mathcal{F}}}$.
If the frequency domain channel estimate ${\bf{\hat H}}_{sr,k}$ is
computed according to (\ref{f_t_relation}), we have
\begin{align}
\label{f_error} & \mathbb{E}\{ {\rm{vec}}([
\Delta{{{\bf{H}}}_{sr,0}} \ \cdots \
\Delta{{\bf{H}}}_{sr,K-1}])\nonumber \\
& \ \ \ \ \ \ \ \ \ \times{\rm{vec}}^{\rm{H}}([
\Delta{{{\bf{H}}}_{sr,0}} \ \cdots \ \Delta{{\bf{H}}}_{sr,K-1}])\}
\nonumber \\
 = &({\boldsymbol {\mathcal{F}}}_{L_1} \otimes
{\bf{I}}_{M_RN_S})\underbrace{({\boldsymbol
\Lambda}_{\rm{channel}}^{-1}\otimes{\bf{I}}_{N_S}+{{\sigma_{n_1}^{-2}}({\bf{D}}^{*}{\bf{D}}^{\rm{T}})})^{-1}}_{\triangleq{\boldsymbol
\Phi}^{sr}}\otimes{\bf{I}}_{M_R}\nonumber \\
& \times ({\boldsymbol {\mathcal{F}}}_{L_1}
\otimes {\bf{I}}_{M_R N_S})^{\rm{H}}K,
\end{align} where $\Delta{\bf{H}}_{sr,k}={\bf{H}}_{sr,k}-{\bf{\hat H}}_{sr,k}$.

In case there is no prior information on ${\bf{R}}_{\rm{channel}}$, we can assign uninformative prior to ${\boldsymbol{\xi}}_{sr}$, that is, $\sigma_{h_0},\sigma_{h_1},\cdots,\sigma_{h_{L-1}}$ approach infinity \cite{Robert01}. In this case,
${\bf{R}}_{\rm{channel}}^{-1}\rightarrow {\bf{0}}$, and then the channel estimator (\ref{estimtaion_MMSE})
and estimation MSE (\ref{MSE_LMMSE}) reduce to that of maximum likelihood (ML) estimation \cite[P.179]{Larsson03}.

Taking the $M_RN_S \times M_RN_S$ block diagonal
elements from (\ref{f_error}) gives
\begin{align}
\label{E_1}& \mathbb{E}\{{\rm{vec}}(\Delta{\bf{{H}}}_{sr,k})
{\rm{vec}}^{\rm{H}}(\Delta{\bf{{H}}}_{sr,k})\} \nonumber \\
=&\left(\sum_{\ell_2=0}^{L_1-1}\sum_{\ell_1=0}^{L_1-1}(e^{-j\frac{2\pi}{K}k(\ell_{1}-\ell_{2})}{\boldsymbol{\Phi}}_{\ell_{1},\ell_{2}}^{sr})\right)\otimes{\bf{I}}_{M_R}.
\end{align} where ${\boldsymbol{\Phi}}_{\ell_{1},\ell_{2}}^{sr}$ is the $N_S \times N_S$ matrix taken
from the following partition of ${\boldsymbol \Phi}^{sr}$
\begin{align}
\label{E_matrix} {\boldsymbol{\Phi}}^{sr}& = \left[
{\begin{array}{*{20}c}
   {\boldsymbol{\Phi}}_{0,0}^{sr} & {\boldsymbol{\Phi}}_{0,1}^{sr}   &  \cdots  &  {\boldsymbol{\Phi}}_{0,L_1-1}^{sr}  \\
    \vdots  &  \cdots  &  \ddots & \vdots   \\
  {\boldsymbol{\Phi}}_{L_1-1,0}^{sr} &  {\boldsymbol{\Phi}}_{L_1-1,1}^{sr} &  \cdots  & {\boldsymbol{\Phi}}_{L_1-1,L_1-1}^{sr} \\
\end{array}} \right].
\end{align}
Furthermore, based on (\ref{E_1}), for an arbitrary square matrix
${\bf{R}}$, it is proved in Appendix~\ref{Lemma} that
\begin{align}
& \mathbb{E}\{\Delta{\bf{H}}_{sr,k}{\bf{R}}
\Delta{\bf{H}}^{\rm{H}}_{sr,k}\} \nonumber \\
&={\rm{Tr}}\left({\bf{R}}{\sum_{\ell_{2}=0}^{L_1-1}\sum_{\ell_{1}=0}^{L_1-1}\left(e^{-j\frac{2\pi}{K}k(\ell_{1}-\ell_{2})}{({\boldsymbol
\Phi}^{sr}_{\ell_{1},\ell_{2}})}^{\rm{T}}\right)
}\right){\bf{I}}_{M_R}. \label{Expectaion_H_sr}
\end{align}

A similar result holds for the second hop. In particular, denoting
the relationship between the true value and estimate of the second
hop channel as
\begin{align}
{\bf{H}}_{rd,k}&={\bf{\hat H}}_{rd,k}+\Delta{\bf{H}}_{rd,k}, \ \ \
k=0, \cdots, K-1,
\end{align} we have the following property
\begin{align}
& \mathbb{E}\{\Delta{\bf{H}}_{rd,k}{\bf{R}}
\Delta{\bf{H}}^{\rm{H}}_{rd,k}\} \nonumber \\
&={\rm{Tr}}\left({\bf{R}}{\sum_{\ell_{1}=0}^{L_2-1}\sum_{\ell_{2}=0}^{L_2-1}\left(e^{-j\frac{2\pi}{K}k(\ell_{1}-\ell_{2})}{({\boldsymbol
\Phi}^{rd}_{\ell_{1},\ell_{2}})}^{\rm{T}}\right)
}\right){\bf{I}}_{M_D}, \label{Expectaion_H_rd}
\end{align} where $L_2$ is the length of the second hop channel in
time domain. Furthermore, as the two channels are estimated independently, $\Delta{\bf{H}}_{sr,k}$ and $\Delta{\bf{H}}_{rd,k}$ are independent.

\section{Transceiver Design Problem Formulation}
\label{problemformulation}

At the destination, a linear equalizer ${\bf{G}}_k$ is adopted for
each subcarrier to detect the transmitted data ${\bf{s}}_k$ (see
Fig.~\ref{fig:1}). The problem is how to design the linear
forwarding matrix matrix ${\bf{F}}_k$ at the relay and the linear
equalizer ${\bf{G}}_k$ at the destination to minimize the MSE of the
received data at the destination:
\begin{align}
\label{equ:MSE_0_1} &{\rm{MSE}}_k({\bf{F}}_k,{\bf{G}}_k)=
{\mathbb{E}} \{ {\rm{Tr}} \left (
({\bf{G}}_k{\bf{y}}_k-{\bf{s}}_k)({\bf{G}}_k{\bf{y}}_k-{\bf{s}}_k)^{\rm{H}}
\right ) \},
\end{align}
where the expectation is taken with respect to ${\bf{s}}_k$, $\Delta
{\bf{H}}_{sr,k}$, $\Delta {\bf{H}}_{rd,k}$, ${\bf{n}}_{1,k}$ and
${\bf{n}}_{2,k}$\footnote{In this paper, the MSE is in fact an
average of the traditional MSE over all possible channel estimation
errors $\Delta {\bf{H}}_{sr,k}$ and $\Delta {\bf{H}}_{rd,k}$.
 When the LMMSE channel estimator is adopted, it is equivalent to the
 conditional MSE corresponding to the partial CSI case defined in \cite{Dietrich07}.}. Since ${\bf{s}}_k$, ${\bf{n}}_{1,k}$ and
${\bf{n}}_{2,k}$ are independent, the MSE expression
(\ref{equ:MSE_0_1}) can be written as
\begin{align}
\label{equ:MSE}
&{\rm{MSE}}_k({{\bf{F}}_k,{\bf{G}}}_k) \nonumber \\
=& {\mathbb{E}} \{ \|({\bf{G}}_k{{\bf{H}}_{rd,k} {\bf{F}}_k
{\bf{H}}_{sr,k}-{\bf{I}}_{N_S})
{\bf{s}}_k} + {\bf{G}}_k{{\bf{H}}_{rd,k} {\bf{F}}_k{\bf{n}}_{1,k} }\nonumber \\
&+ {\bf{G}}_k{\bf{n}}_{2,k} \|^2 \}\nonumber \\
=& {\mathbb{E}}_{\Delta{\bf{H}}_{sr,k},\Delta{\bf{H}}_{rd,k}} \{
{\rm{Tr}}( ({\bf{G}}_k{\bf{H}}_{rd,k}{\bf{F}}_k
{\bf{H}}_{sr,k}-{\bf{I}}_{N_S}){\bf{R}}_{{\bf{s}}_k} \nonumber \\
& \times ({\bf{G}}_k{\bf{H}}_{rd,k}{\bf{F}}_k{\bf{H}}_{sr,k}-{\bf{I}}_{N_S})^{\rm{H}})
  \}\nonumber \\
&+{\mathbb{E}}_{\Delta{\bf{H}}_{rd,k}} \{ {\rm{Tr}} \left(
({\bf{G}}_k{\bf{H}}_{rd,k}{\bf{F}}_k)
{\bf{R}}_{n_{1,k}}({\bf{G}}_k{\bf{H}}_{rd,k}{\bf{F}}_k)^{\rm{H}} \right) \}\nonumber \\
& + {\rm{Tr}}({\bf{G}}_k{\bf{R}}_{n_{2,k}}{\bf{G}}_k^{\rm{H}})  \nonumber \\
=&{\mathbb{E}}_{\Delta{\bf{H}}_{sr,k},\Delta{\bf{H}}_{rd,k}} \{
{\rm{Tr}}  (
({\bf{G}}_k{\bf{H}}_{rd,k}{\bf{F}}_k{\bf{H}}_{sr,k}){\bf{R}}_{{\bf{s}}_k}\nonumber \\
&\times({\bf{G}}_k{\bf{H}}_{rd,k}
{\bf{F}}_k{\bf{H}}_{sr,k})^{\rm{H}}) \}\nonumber \\
&+{\rm{Tr}} ( {\bf{G}}_k{\mathbb{E}}_{\Delta{\bf{H}}_{rd,k}} \{
{\bf{H}}_{rd,k}{\bf{F}}_k
{\bf{R}}_{n_{1,k}}{\bf{F}}_k^{\rm{H}}{\bf{H}}_{rd,k}^{\rm{H}}
\}{\bf{G}}_k
^{\rm{H}})\nonumber \\
&-{\rm{Tr}} ( ({\bf{G}}_k{\bf{\hat
H}}_{rd,k}{\bf{F}}_k{\bf{\hat H}}_{sr,k}{\bf{R}}_{s,k})^{\rm{H}}
)\nonumber \\
 &-{\rm{Tr}} ({\bf{G}}_k{\bf{\hat H}}_{rd,k}{\bf{F}}_k{\bf{\hat
H}}_{sr,k}{\bf{R}}_{s,k})\nonumber \\
& +{\rm{Tr}} ( {\bf{R}}_{{\bf{s}}_k}
)+{\rm{Tr}}({\bf{G}}_k{\bf{R}}_{n_{2,k}}{\bf{G}}_k^{\rm{H}}).
\end{align}
Because $\Delta{\bf{H}}_{sr,k}$ and $\Delta{\bf{H}}_{rd,k}$ are
independent, the first term of $\rm{MSE}_k$ is
\begin{align}
\label{equ:MSE_1_a} &
{\mathbb{E}}_{\Delta{\bf{H}}_{sr,k},\Delta{\bf{H}}_{rd,k}} \{
{\rm{Tr}} (
({\bf{G}}_k{\bf{H}}_{rd,k}{\bf{F}}_k{\bf{H}}_{sr,k})
{\bf{R}}_{{\bf{s}}_k} \nonumber \\
& \ \ \ \ \ \ \ \ \ \ \ \ \ \ \ \ \ \ \ \ \ \ \ \ \ \ \ \ \ \ \times ({\bf{G}}_k{\bf{H}}_{rd,k}{\bf{F}}_k{\bf{H}}_{sr,k})^{\rm{H}}
 ) \} \nonumber \\
&= {\rm{Tr}}( {\bf{G}}_k{\mathbb{E}}_{{\Delta\bf{H}}_{rd,k}}
\{
{\bf{H}}_{rd,k}{\bf{F}}_k\nonumber \\
& \ \ \ \ \ \ \ \ \ \ \ \times {\mathbb{E}}_{{\Delta\bf{H}}_{sr,k}} \{
{\bf{H}}_{sr,k}{\bf{R}}_{{\bf{s}}_k}{\bf{H}}_{sr,k}^{\rm{H}}
\}{\bf{F}}_k^{\rm{H}} {\bf{H}}_{rd,k}^{\rm{H}}\}
{\bf{G}}_k^{\rm{H}}).
\end{align} For the inner expectation, the following equation holds
\begin{align}
\label{equ:iteg} &{\mathbb{E}}_{\Delta{\bf{H}}_{sr,k} } \{
{\bf{H}}_{sr,k}{\bf{R}}_{{\bf{s}}_k}{\bf{H}}_{sr,k}^{\rm{H}} \}\nonumber \\
&={\mathbb{E}}_{\Delta{\bf{H}}_{sr,k} } \{ ({\bf{\hat
H}}_{sr,k}+\Delta
{\bf{H}}_{sr,k} ){\bf{R}}_{{\bf{s}}_k}({\bf{\hat H}}_{sr,k}+\Delta{\bf{H}}_{sr,k})^{\rm{H}} \} \nonumber \\
&={\rm{Tr}}({\bf{R}}_{{\bf{s}}_k} {\boldsymbol{\Psi}}
_{sr,k}){\bf{I}}_{M_R} + {\bf{\hat
H}}_{sr,k}{\bf{R}}_{{\bf{s}}_k}{\bf{\hat
H}}_{sr,k}^{\rm{H}}\triangleq {\boldsymbol\Pi}_{k},
\end{align}where based on (\ref{Expectaion_H_sr}) the matrix ${\boldsymbol \Psi}_{sr,k}$ is defined
as
\begin{align}
\label{Psi_sr} {\boldsymbol
\Psi}_{sr,k}=\sum_{\ell_{1}=0}^{L_1-1}\sum_{\ell_{2}=0}^{L_1-1}\left(e^{-j\frac{2\pi}{K}k(\ell_{1}-\ell_{2})}{({\boldsymbol
\Phi}^{sr}_{\ell_{1},\ell_{2}})}^{\rm{T}}\right).
\end{align}

Applying (\ref{equ:iteg}) and the corresponding result for $\Delta
{\bf{H}}_{rd,k}$ to (\ref{equ:MSE_1_a}), the first term of
${\rm{MSE}}_k$ becomes
\begin{align}
\label{equ:MSE_1} &{\rm{Tr}}  ( {\bf{G}}_k{\mathbb{E}}_{\Delta
{\bf{H}}_{rd,k}} \{
{\bf{H}}_{rd,k}{\bf{F}}_k \nonumber \\
& \ \ \ \ \ \ \ \ \ \ \ \ \ \times {\mathbb{E}}_{\Delta {\bf{H}}_{sr,k}} \{
{\bf{H}}_{sr,k}{\bf{R}}_{{\bf{s}}_k}{\bf{H}}_{sr,k}^{\rm{H}}
\}{\bf{F}}_k^{\rm{H}}
{\bf{H}}_{rd,k}^{\rm{H}}\} {\bf{G}}_k^{\rm{H}} ) \nonumber \\
=&{\rm{Tr}}( {\bf{G}}_k{\bf{G}}_k^{\rm{H}}
   ){\rm{Tr}}({\bf{F}}_k{\boldsymbol\Pi}_{k}
{\bf{F}}_k^{\rm{H}} {\boldsymbol \Psi}_{rd,k}
   ) \nonumber \\
   & + {\rm{Tr}}( {\bf{G}}_k{\bf{\hat H}}_{rd,k}
   {\bf{F}}_k{\boldsymbol\Pi}_{k}
   {\bf{F}}_k^{\rm{H}} {\bf{\hat H}}_{rd,k}^{\rm{H}}{\bf{G}}_k^{\rm{H}}
   ),
\end{align} where the matrix ${\boldsymbol \Psi}_{rd,k}$ is defined
as
\begin{align}
\label{Psi_rd} {\boldsymbol
\Psi}_{rd,k}=\sum_{\ell_{1}=0}^{L_2-1}\sum_{\ell_{2}=0}^{L_2-1}\left(e^{-j\frac{2\pi}{K}k(\ell_{1}-\ell_{2})}{({\boldsymbol
\Phi}^{rd}_{\ell_{1},\ell_{2}})}^{\rm{T}}\right).
\end{align}

Similarly, the second term of ${\rm{MSE}}_k$ in (\ref{equ:MSE}) can
be simplified as
\begin{align}
\label{equ:MSE_2} &{\rm{Tr}} \left(
{\bf{G}}_k{\mathbb{E}}_{\Delta{\bf{H}}_{rd,k}} \{
{\bf{H}}_{rd,k}{\bf{F}}_k
{\bf{R}}_{n_{1},k}{\bf{F}}_k^{\rm{H}}{\bf{H}}_{rd,k}^{\rm{H}}
\}{\bf{G}}_k
^{\rm{H}} \right) \nonumber \\
=&{\rm{Tr}}( {\bf{G}}_k{\bf{G}}_k^{\rm{H}}){\rm{Tr}}({\bf{F}}_k{\bf{R}}_{n_{1,k}}
{\bf{F}}_k^{\rm{H}}{\boldsymbol{\Psi}} _{rd,k} ) \nonumber \\ &+
{\rm{Tr}}( {\bf{G}}_k{\bf{\hat H}}_{rd,k} {\bf{F}}_k {\bf{R}}_{n_{1},k}
{\bf{F}}_k^{\rm{H}} {\bf{\hat
H}}_{rd,k}^{\rm{H}}{\bf{G}}_k^{\rm{H}}).
\end{align}
Based on (\ref{equ:MSE_1}) and (\ref{equ:MSE_2}), the
${\rm{MSE}}_k$ (\ref{equ:MSE}) equals to
\begin{align}
\label{MSE}& {\rm{MSE}}_k({{\bf{F}}_k,{\bf{G}}}_k)\nonumber \\
=&\ {\rm{Tr}}
( {\bf{G}}_k({\bf{\hat
H}}_{rd,k}{\bf{F}}_k{\bf{R}}_{{\bf{x}}_k}{\bf{F}}_k^{\rm{H}}
{\bf{\hat H}}_{rd,k}^{\rm{H}}+{\bf{K}}_k){\bf{G}}_k^{\rm{H}}
) \nonumber \\
&-{\rm{Tr}} ( {\bf{R}}_{{\bf{s}}_k}{\bf{\hat
H}}_{sr,k}^{\rm{H}}{\bf{F}}_k^{\rm{H}}{\bf{\hat
H}}_{rd,k}^{\rm{H}}{\bf{G}}_k^{\rm{H}}  )-{\rm{Tr}}( {\bf{G}}_k{\bf{\hat
H}}_{rd,k}{\bf{F}}_k{\bf{\hat H}}_{sr,k}{\bf{R}}_{{\bf{s}}_k}
)\nonumber
\\
& +{\rm{Tr}} ( {\bf{R}}_{{\bf{s}}_k} )
\end{align} where
\begin{align}
{\bf{R}}_{{\bf{x}}_k}&=
{\boldsymbol\Pi}_{k}+\sigma_{n_1}^2{\bf{I}}_{M_R}
\label{R_x}\\
{\bf{K}}_k&=
{({\rm{Tr}}({\bf{F}}_k{\bf{R}}_{{\bf{x}}_k}{\bf{F}}_k^{\rm{H}}
{\boldsymbol\Psi} _{rd,k})+\sigma_{n_2}^2)}{\bf{I}}_{M_D} \nonumber \\
 &\triangleq \eta_k {\bf{I}}_{M_D}. \label{K}
\end{align} Notice that the matrix ${\bf{R}}_{{\bf{x}}_k}$ is the correlation
matrix of the receive signal ${\bf{x}}_k$ on the $k^{\rm{th}}$
subcarrier at the relay.

Subject to the transmit power constraint at the relay, the joint
design of relay forwarding matrix and destination equalizer that
minimizes the total MSE of the output data at the destination can be
formulated as the following optimization problem
\begin{align}
\label{MSE_Opt}
& \min \limits_{{\bf{F}}_k,{\bf{G}}_k}\ \ \sum_k {\rm{MSE}}_k({\bf{F}}_k,{\bf{G}}_k) \nonumber \\
& \ {\rm{s.t.}} \ \ \ \
\sum_k{\rm{Tr}}({\bf{F}}_k{\bf{R}}_{{\bf{x}}_k}{\bf{F}}_k^{\rm{H}})
\le P_r.
\end{align}

\textbf{\textsl{Remark 1:}} In this paper, the relay estimates the
source-relay channel and the destination estimates the
relay-destination channel. The forwarding matrix ${\bf{F}}_k$ and
equalizer ${\bf{G}}_k$ are designed at the relay. Therefore, the
estimated second hop CSI should be fed back from destination to
relay. However, when channel is varying slowly, and the channel
estimation feedback occurs infrequently, the errors in feedback can
be negligible.

\section{Proposed Closed-Form Solution for ${\bf{G}}_k$'s and ${\bf{F}}_k$'s}
\label{closedformsolution}

In this section, we will derive a closed-form solution for the
optimization problem (\ref{MSE_Opt}). In order to facilitate the
analysis, the optimization problem (\ref{MSE_Opt}) is rewritten as
\begin{align}
\label{MSE_Opt_1}
& \min \limits_{{\bf{F}}_k,{\bf{G}}_k,{P}_{r,k}}\ \ \sum_k{\rm{MSE}}_k({\bf{F}}_k,{\bf{G}}_k) \nonumber \\
& \ {\rm{s.t.}} \ \ \ \ \ \ \ \ \
{\rm{Tr}}({\bf{F}}_k{\bf{R}}_{{\bf{x}}_k}{\bf{F}}_k^{\rm{H}}) \le
P_{r,k},
\ \ \ \ k=0,\cdots,K-1  \nonumber \\
& \ \ \ \ \ \ \ \ \ \ \ \ \ \ \sum_k P_{r,k} \le P_r,
\end{align} with the physical meaning of $P_{r,k}$ being the maximum allocated power over the
$k^{\rm{th}}$ subcarrier.

The Lagrangian function of the
optimization problem (\ref{MSE_Opt_1}) is
\begin{align}
\label{L_function} &\mathcal{L}({\bf{F}}_k,{\bf{G}}_k,{P}_{r,k})
=\sum_{k}{\rm{MSE}}_{k}({\bf{F}}_k,{\bf{G}}_k)+
\sum_{k}\gamma_{k}({\rm{Tr}}
({\bf{F}}_k{\bf{R}}_{{\bf{x}}_k}{\bf{F}}_k^{\rm{H}})\nonumber \\
&-P_{r,k})+\rho(\sum_k
P_{r,k}-P_r)
\end{align} where the positive scalars $\gamma_k$ and $\rho$ are the Lagrange multipliers.
Differentiating (\ref{L_function}) with respect to ${\bf{F}}_k$,
${\bf{G}}_k$ and $P_{r,k}$, and setting the corresponding results to
zero, the Karush-Kuhn-Tucker (KKT) conditions of the optimization
problem (\ref{MSE_Opt_1}) are given by \cite{Boyd04}
\begin{subequations}
\begin{align}
&{\bf{G}}_k({\bf{\hat
H}}_{rd,k}{\bf{F}}_k{\bf{R}}_{{\bf{x}}_k}{\bf{F}}_k^{\rm{H}}
{\bf{\hat H}}_{rd,k}^{\rm{H}}+{\bf{K}}_{k})=
{\bf{R}}_{{\bf{s}}_k}({\bf{\hat
H}}_{rd,k}{\bf{F}}_k{\bf{\hat H}}_{sr,k})^{\rm{H}}, \label{equ:L_1}\\
 & {\bf{\hat
H}}_{rd,k}^{\rm{H}}{\bf{G}}_k^{\rm{H}}{\bf{G}}_k{\bf{\hat
H}}_{rd,k}{\bf{F}}_k{\bf{R}}_{{\bf{x}}_k}+({\rm{Tr}}({\bf{G}}_k{\bf{G}}_k^{\rm{H}})
{\boldsymbol\Psi} _{rd,k}
+\gamma_k{\bf{I}}_{N_R})\nonumber \\&\times{\bf{F}}_k{\bf{R}}_{{\bf{x}}_k}= \left
({\bf{\hat H}}_{sr,k}{\bf{R}}_{{\bf{s}}_k}{\bf{G}}_k{\bf{\hat
H}}_{rd,k}\right)^{\rm{H}}, \label{equ:L_2} \\
& \gamma_k({\rm{Tr}}({\bf{F}}_k{\bf{R}}_{{\bf{x}}_k}{\bf{F}}_k^{\rm{H}})-P_{r,k})=0,\label{equ:L_3} \\
& \gamma_k\ge 0, \ \ \ \  k=0,\cdots,K-1, \label{equ:L_4}  \\
&  \rho(\sum_k P_{r,k}-P_r)=0, \label{equ:L_5} \\
&  \gamma_0=\gamma_1=\cdots=\gamma_{K-1}=\rho, \label{equ:L_6}\\
& {\rm{Tr}}({\bf{F}}_k{\bf{R}}_{{\bf{x}}_k}{\bf{F}}_k^{\rm{H}})\le
P_{r,k}, \label{equ:L_7} \\
& \sum_k P_{r,k} \le P_r. \label{equ:L_8}
\end{align}
\end{subequations}

It is obvious that the objective function and constraints of
(\ref{MSE_Opt_1}) are continuously differentiable. Furthermore, it
is easy to see that solutions of the optimization problem
(\ref{MSE_Opt_1}) satisfy the regularity condition, i.e., Abadie
constraint qualification (ACQ), because linear independence
constraint qualification (LICQ) can be proved \cite{Bertsekas03}.
Based on these facts, the KKT conditions are the necessary
conditions.\footnote{Notice that the solution
${\bf{F}}_0=\cdots={\bf{F}}_{K-1}={\bf{0}}$ and
${\bf{G}}_0=\cdots={\bf{G}}_{K-1}={\bf{0}}$ also satisfies the KKT
conditions, but this solution is meaningless as no signal can be
transmitted \cite{Sampath01}.} From KKT conditions, we can derive
the following two useful properties which can help us to find the
optimal solution.


\noindent \textbf{\textsl{\underline {Property 1:}}}  It is proved in
Appendix~\ref{Lemma_1} that for any ${\bf{F}}_k$ satisfying the KKT
conditions (\ref{equ:L_1})-(\ref{equ:L_5}), the power
constraints (\ref{equ:L_7}) and (\ref{equ:L_8}) must occur on the boundaries
\begin{align}
{\rm{Tr}}({\bf{F}}_{k}{\bf{R}}_{{\bf{x}}_k}{\bf{F}}_{k}^{\rm{H}})&=P_{r,k},
\label{sum_F} \\
\sum_k P_{r,k}&=P_r \label{power_sum}.
\end{align} Furthermore, the corresponding ${\bf{G}}_k$ satisfies
\begin{align}
\label{G_P_a}
{\rm{Tr}}({\bf{G}}_k{\bf{G}}_k^{\rm{H}})={\gamma_kP_{r,k}}/{\sigma_{n_2}^2}.
\end{align}

$ \ $

\noindent \textbf{\textsl{\underline{Property 2:}}} Define the
matrices ${\bf{U}}_{{\bf{T}}_k}$, ${\bf{V}}_{{\bf{T}}_k}$ ,
${\boldsymbol { \Lambda}}_{{\bf{T}}_k}$,
${\bf{U}}_{{\boldsymbol{\Theta}}_k}$, and ${\boldsymbol {
\Lambda}}_{{\boldsymbol{\Theta}}_k}$ based on eigenvalue
decomposition (EVD) and singular value decomposition (SVD) as
\begin{align}
& (P_{r,k} {\boldsymbol\Psi}
_{rd,k}+\sigma_{n_2}^2{\bf{I}}_{N_R})^{-\frac{\rm{H}}{2}}{\bf{\hat
H}}_{rd,k}^{\rm{H}}\nonumber \\
& \ \ \ \ \ \ \ \ \  \times \underbrace{{\bf{\hat H}}_{rd,k}(P_{r,k}
{\boldsymbol\Psi}
_{rd,k}+\sigma_{n_2}^2{\bf{I}}_{N_R})^{-\frac{1}{2}}}_{\triangleq{\boldsymbol
\Theta}_k}= {\bf{U}}_{{\boldsymbol{\Theta}}_k}{\boldsymbol
{\Lambda}}_{{\boldsymbol{\Theta}}_k}{\bf{U}}_{{\boldsymbol{\Theta}}_k}^{\rm{H}}\label{case_2},
\\
& {\bf{R}}_{{\bf{x}},k}^{-\frac{1}{2}}{\bf{\hat
H}}_{sr,k}{\bf{R}}_{{\bf{s}}_k} ={\bf{U}}_{{\bf{T}}_k}{\boldsymbol
\Lambda}_{{\bf{T}}_k}{\bf{V}}_{{\bf{T}}_k}^{\rm{H}}\label{case_3},
\end{align} with elements of the diagonal matrix ${\boldsymbol
{\Lambda}}_{{\bf{T}}_k}$ and ${\boldsymbol
{\Lambda}}_{{\boldsymbol{\Theta}}_k}$ arranged in decreasing order.
Then with KKT conditions (\ref{equ:L_1}) and (\ref{equ:L_2}), it is
proved in Appendix~\ref{Lemma2} that the optimal forwarding matrix
${\bf{F}}_k$ and equalizer ${\bf{G}}_k$ must be in
the form
\begin{align}
&{\bf{F}}_{k}=(P_{r,k} {\boldsymbol\Psi}
_{rd,k}+\sigma_{n_2}^2{\bf{I}}_{N_R})^{-\frac{1}{2}}{\bf{U}}_{{\boldsymbol{\Theta}}_k,q_k}{\boldsymbol
A}_{{\bf{F}}_{k}}{\bf{U}}_{{\bf{T}}_k,p_k}^{\rm{H}} {\bf{R}}_{{\bf{x}}_k}^{-\frac{1}{2}} \label{F_P_2}, \\
&{\bf{G}}_{k}={\bf{V}}_{{\bf{T}}_k,p_k}{\boldsymbol
A}_{{\bf{G}}_k}{\bf{U}}_{{\boldsymbol{\Theta}}_k,q_k}^{\rm{H}}(P_{r,k}
{\boldsymbol\Psi}
_{rd,k}+\sigma_{n_2}^2{\bf{I}}_{N_R})^{-\frac{\rm{H}}{2}} {\bf{\hat
H}}_{rd,k}^{\rm{H}} \label{G_P_2},
\end{align} where ${\boldsymbol A}_{{\bf{F}}_k}$ and ${\boldsymbol A}_{{\bf{G}}_k}$ are to be determined. The matrix
${\bf{U}}_{{\bf{T}}_k,p_k}$ and
${\bf{V}}_{{\bf{T}}_k,p_k}$ are the first $p_k$ columns
of ${\bf{U}}_{{\bf{T}}_k}$ and ${\bf{V}}_{{\bf{T}}_k}$, respectively, and $p_k={\rm{Rank}}({\boldsymbol {\Lambda}}_{{\bf{T}}_k})$. Similarly,  ${\bf{U}}_{{\boldsymbol{\Theta}}_k,q_k}$ is the first $q_k$ columns
of ${\bf{U}}_{{\boldsymbol{\Theta}}_k}$, and $q_k={\rm{Rank}}({\boldsymbol {
\Lambda}}_{{\boldsymbol{\Theta}}_k})$.

$ \ $

Right multiplying both sides
of (\ref{equ:L_1}) with ${\bf{G}}_k^{\rm{H}}$ and left multiplying
both sides of (\ref{equ:L_2}) with ${\bf{F}}_k^{\rm{H}}$, and making
use of (\ref{F_P_2}) and (\ref{G_P_2}), the first two KKT conditions become
\begin{align}
&{\boldsymbol {A}}_{{\bf{G}}_k}{\boldsymbol {\bar
\Lambda}}_{{\boldsymbol{\Theta}}_k}{\boldsymbol
{A}}_{{\bf{F}}_k}{\boldsymbol {A}}_{{\bf{F}}_k}^{\rm{H}}{\boldsymbol
{\bar \Lambda}}_{{\boldsymbol{\Theta}}_k}{\boldsymbol
{A}}_{{\bf{G}}_k}^{\rm{H}}+\eta_k{\boldsymbol
{A}}_{{\bf{G}}_k}{\boldsymbol {\bar
\Lambda}}_{{\boldsymbol{\Theta}}_k}{\boldsymbol
{A}}_{{\bf{G}}_k}^{\rm{H}}\nonumber \\&=({\boldsymbol
{A}}_{{\bf{G}}_k}{\boldsymbol {\bar
\Lambda}}_{{\boldsymbol{\Theta}}_k}{\boldsymbol
{A}}_{{\bf{F}}_k}{\boldsymbol {\bar
\Lambda}}_{{\bf{T}}_k})^{\rm{H}}\label{F_3}, \\
& {\boldsymbol {A}}_{{\bf{F}}_k}^{\rm{H}}{\boldsymbol {\bar
\Lambda}}_{{\boldsymbol{\Theta}}_k}{\boldsymbol
{A}}_{{\bf{G}}_k}^{\rm{H}}{\boldsymbol {A}}_{{\bf{G}}_k}{\boldsymbol
{\bar \Lambda}}_{{\boldsymbol{\Theta}}_k}{\boldsymbol
{A}}_{{\bf{F}}_k}+\frac{\gamma_k}{\sigma_{n_2}^2}{\boldsymbol
{A}}_{{\bf{F}}_k}^{\rm{H}}{\boldsymbol
{A}}_{{\bf{F}}_k}\nonumber \\
&=({\boldsymbol {\bar
\Lambda}}_{{\bf{T}}_k}{\boldsymbol {A}}_{{\bf{G}}_k}{\boldsymbol
{\bar \Lambda}}_{{\boldsymbol \Theta}_k}{\boldsymbol
{A}}_{{\bf{F}}_k})^{\rm{H}}\label{G_3},
\end{align} where the matrix ${\boldsymbol {\bar
\Lambda}}_{{\boldsymbol{\Theta}}_k}$ is the $q_k\times q_k$
principal submatrix of ${\boldsymbol {
\Lambda}}_{{\boldsymbol{\Theta}}_k}$. Similarly, ${\boldsymbol {\bar
\Lambda}}_{{\bf{T}}_k}$ is the $p_k\times p_k$ principal submatrix
of ${\boldsymbol { \Lambda}}_{{\bf{T}}_k}$. In this paper, we
consider AF MIMO-OFDM relay systems, the matrices ${\boldsymbol
{A}}_{{\bf{F}}_k}$ and ${\boldsymbol{A}}_{{\bf{G}}_k}$ can be of
arbitrary dimension instead of the square matrices considered in
point-to-point systems \cite{Sampath01}, \cite{Ding09}. Then, the
solutions satisfying KKT conditions and obtained by solving
(\ref{F_3}) and (\ref{G_3}) are not unique. To identify the optimal
solution, we need an additional information which is presented in
the following \textbf{\textsl{{Property 3}}}.

\noindent \textbf{\textsl{\underline {Property 3:}}} Putting the results of \textbf{\textsl{{Property 1}}} and \textbf{\textsl{{Property 2}}} into the optimization problem (\ref{MSE_Opt_1}), based on majorization theory, it is proved in Appendix~\ref{Lemma_3} that the optimal ${\boldsymbol{A}}_{{\bf{F}}_k}$ and ${\boldsymbol{A}}_{{\bf{G}}_k}$ have the following diagonal structure
\begin{align}
{\boldsymbol{A}}_{{\bf{F}}_k,{\rm{opt}}}&=\left[
{\begin{array}{*{20}c}
  {\boldsymbol{
\Lambda}}_{{\bf{F}}_k,{\rm{opt}}} & {\bf{0}}_{N_k,p_k-N_k} \\
   {\bf{0}}_{q_k-N_k,N_k} &  {\bf{0}}_{q_k-N_k,p_k-N_k}\\
\end{array}} \right], \label{F_A}\\
{\boldsymbol{A}}_{{\bf{G}}_k,{\rm{opt}}}&=\left[
{\begin{array}{*{20}c}
  {\boldsymbol{
\Lambda}}_{{\bf{G}}_k,{\rm{opt}}} & {\bf{0}}_{N_k,q_k-N_k} \\
   {\bf{0}}_{p_k-N_k,N_k} &  {\bf{0}}_{p_k-N_k,q_k-N_k}\\
\end{array}} \right], \label{G_A}
\end{align} where ${\boldsymbol{\Lambda}}_{{\bf{ F}}_k, {\rm{opt}}}$ and ${\boldsymbol{\Lambda}}_{{\bf{ G}}_k, {\rm{opt}}}$  are two $N_k \times N_k$ diagonal
matrices to be determined, and $N_k={{\min}}(p_k,q_k)$. Notice that
Property 3 is obtained by applying majorization theory to the
original optimization problem.  It is also a necessary condition for
the optimal solution, and contains different information from that
of Property 2.

$ \ $

Combining Property 2 and Property 3, and following the argument in
\cite{Sampath01}, it can be concluded that the optimal solution of
of ${\boldsymbol {A}}_{{\bf{F}}_k}$ and
${\boldsymbol{A}}_{{\bf{G}}_k}$ is unique. Now, substituting
(\ref{F_A}) and (\ref{G_A}) into (\ref{F_3}) and (\ref{G_3}), and
noticing that all matrices are diagonal, ${\boldsymbol
{\Lambda}}_{{\bf{F}}_k,{\rm{opt}}}$ and ${\boldsymbol
{\Lambda}}_{{\bf{G}}_k,{\rm{opt}}}$ can be easily solved to be
\begin{align}
{\boldsymbol {\Lambda}}_{{\bf{F}}_k,{\rm{opt}}}&=
\left[\left(\sqrt{\frac{\sigma_{n_2}^2
\eta_k}{\gamma_k}}{\boldsymbol {\tilde
\Lambda}}_{{\boldsymbol{\Theta}}_k}^{-\frac{1}{2}}{\boldsymbol
{\tilde \Lambda}}_{{\bf{T}}_k}-\eta_k{\boldsymbol {\tilde
\Lambda}}_{{\boldsymbol{\Theta}}_k}^{-1}\right)^{+}\right]^{\frac{1}{2}}, \label{F_opt_aa}
\\
 {\boldsymbol {\Lambda}}_{{\bf{G}_k},{\rm{opt}}}&=
\left[\left(\sqrt{\frac{\gamma_k}{\eta_k\sigma_{n_2}^2}}{\boldsymbol
{\tilde
\Lambda}}_{{\boldsymbol{\Theta}}_k}^{-\frac{1}{2}}{\boldsymbol
{\tilde
\Lambda}}_{{\bf{T}}_k}-\frac{\gamma_k}{\sigma_{n_2}^2}{\boldsymbol
{\tilde
\Lambda}}_{{\boldsymbol{\Theta}}_k}^{-1}\right)^{+}\right]^{\frac{1}{2}}{\boldsymbol
{\tilde \Lambda}}_{{\boldsymbol{\Theta}}_k}^{-\frac{1}{2}}, \label{G_opt_aa}
\end{align} where the matrices
${\boldsymbol {\tilde \Lambda}}_{{\bf{T}}_k}$ and ${\boldsymbol
{\tilde \Lambda}}_{{\boldsymbol{\Theta}}_k}$ are the principal
sub-matrices of
 ${\boldsymbol {
\Lambda}}_{{\bf{T}}_k}$ and ${\boldsymbol {
\Lambda}}_{{\boldsymbol{\Theta}}_k}$ with dimension $N_k \times
N_k$, and $
N_k={\rm{min}}\{{\rm{rank}}({\boldsymbol{\Lambda}}_{{\boldsymbol{\Theta}}_k}),
{\rm{rank}}({\boldsymbol{\Lambda}}_{{\bf{T}}_k}) \}$. The matrices
${\bf{U}}_{{\bf{T}}_k,N_k}$, ${\bf{V}}_{{\bf{T}}_k,N_k}$ and
${\bf{U}}_{{\boldsymbol{\Theta}}_k,N_k}$ are the first $N_k$ columns
of ${\bf{U}}_{{\bf{T}}_k}$, ${\bf{V}}_{{\bf{T}}_k}$ and
${\bf{U}}_{{\boldsymbol{\Theta}}_k}$, respectively. From
(\ref{F_opt_aa}) and (\ref{G_opt_aa}), it can be seen that the
optimal solutions are variants of water-filling solution.
Furthermore, the eigen channels of two hops are paired based on the
best-to-best criterion at the relay.

In the general solution (\ref{F_opt_aa})-(\ref{G_opt_aa}),
$P_{r,k}$, $\eta_k$ and $\gamma_k$ are unknown. However notice that
from (\ref{sum_F}) and (\ref{G_P_a}) in \textbf{\textsl{{Property
1}}}, the optimal forwarding matrix and equalizer should
simultaneously satisfy
\begin{align}
&{\rm{Tr}}({\bf{F}}_{k,{\rm{opt}}}{\bf{R}}_{{\bf{x}}_k}{\bf{F}}_{k,{\rm{opt}}}^{\rm{H}})=P_{r,k}, \label{equ:F_X} \\
&{\rm{Tr}}({\bf{G}}_{k,{\rm{opt}}}{\bf{G}}_{k,{\rm{opt}}}^{\rm{H}})=\gamma_k
P_{r,k}/\sigma^2_{n_2} \label{equ:G_X}.
\end{align}
Substituting (\ref{F_A})-(\ref{G_opt_aa}) into (\ref{equ:F_X})
and (\ref{equ:G_X}), it can be straightforwardly shown that $\eta_k$ and
$\gamma_k$ can be expressed as functions of $P_{r,k}$
\begin{align}
\eta_k&=\frac{b_{3,k}P_{r,k}}{P_{r,k}b_{1,k}+b_{1,k}b_{4,k}-b_{2,k}b_{3,k}}, \label{eta_k} \\
\gamma_k&=\frac{b_{3,k}\sigma_{n_2}^2(P_{r,k}b_{1,k}+b_{1,k}b_{4,k}-b_{2,k}b_{3,k})}{(P_{r,k}+b_{4,k})^2P_{r,k}},
\label{gamma_k}
\end{align} where $b_{1,k}$, $b_{2,k}$, $b_{3,k}$ and $b_{4,k}$ are defined as
\begin{subequations}
\label{auxiliary_b}
\begin{align}
b_{1,k}\triangleq&{\rm{Tr}}({\bf{U}}_{{\boldsymbol
\Theta}_k,N_k}^{\rm{H}}(P_{r,k} {\boldsymbol\Psi}
_{rd,k}+\sigma_{n_2}^2{\bf{I}}_{N_R})^{-1}\nonumber \\
& \times {\bf{U}}_{{\boldsymbol
\Theta}_k,N_k}{\boldsymbol {\tilde
\Lambda}}_{{\bf{T}}_k}{\boldsymbol {\tilde
\Lambda}}_{{\boldsymbol{\Theta}}_k}^{-\frac{1}{2}}{\boldsymbol \Lambda}_{{\bf{I}},k}), \\
b_{2,k}\triangleq &
{\rm{Tr}}({\bf{U}}_{{\boldsymbol{\Theta}}_k,N_k}^{\rm{H}}(P_{r,k}
{\boldsymbol\Psi}
_{rd,k}+\sigma_{n_2}^2{\bf{I}}_{N_R})^{-1}{\bf{U}}_{{\boldsymbol{\Theta}}_k,N_k}{\boldsymbol
{\tilde \Lambda}}_{{\boldsymbol{\Theta}}_k}^{-1}{\boldsymbol \Lambda}_{{\bf{I}},k}),\\
b_{3,k}\triangleq &{\rm{Tr}}({\boldsymbol {\tilde
\Lambda}}_{{\bf{T}}_k}{\boldsymbol {\tilde
\Lambda}}_{{\boldsymbol{\Theta}}_k}^{-\frac{1}{2}}{\boldsymbol \Lambda}_{{\bf{I}},k}),\\
b_{4,k}\triangleq &{\rm{Tr}}({\boldsymbol {\tilde
\Lambda}}_{{\boldsymbol{\Theta}}_k}^{-1}{\boldsymbol
\Lambda}_{{\bf{I}},k}),
\end{align}
\end{subequations} and ${\boldsymbol
\Lambda}_{{\bf{I}},k}$ is a diagonal selection matrix with diagonal
elements being 1 or 0, and serves to replace the operation `+'. Combining all the results in this section, we have the following summary.

\noindent \textbf{\textsl{\underline {Summary:}}} The optimal
forwarding matrix ${\bf{F}}_{k,\rm{opt}}$ and equalizer
${\bf{G}}_{k,\rm{opt}}$ are

\begin{align}
{\bf{F}}_{k,{\rm{opt}}}=&(P_{r,k} {\boldsymbol\Psi}
_{rd,k}+\sigma_{n_2}^2{\bf{I}}_{N_R})^{-\frac{1}{2}}{\bf{U}}_{{\boldsymbol{\Theta}}_k,N_k}{\boldsymbol
\Lambda}_{{\bf{F}}_{k},{\rm{opt}}}\nonumber \\
& {\bf{U}}_{{\bf{T}}_k,N_k}^{\rm{H}} {\bf{R}}_{{\bf{x}}_k}^{-\frac{1}{2}} \label{F_opt}, \\
{\bf{G}}_{k,{\rm{opt}}}=&{\bf{V}}_{{\bf{T}}_k,N_k}{\boldsymbol
\Lambda}_{{\bf{G}}_k,{\rm{opt}}}{\bf{U}}_{{\boldsymbol{\Theta}}_k,N_k}^{\rm{H}}(P_{r,k}
{\boldsymbol\Psi}
_{rd,k}+\sigma_{n_2}^2{\bf{I}}_{N_R})^{-\frac{\rm{H}}{2}} \nonumber \\
&{\bf{\hat
H}}_{rd,k}^{\rm{H}} \label{G_opt},
\end{align}
where
\begin{align}
{\boldsymbol {\Lambda}}_{{\bf{F}}_k,{\rm{opt}}}&=
\left[\left(\sqrt{\frac{\sigma_{n_2}^2
\eta_k}{\gamma_k}}{\boldsymbol {\tilde
\Lambda}}_{{\boldsymbol{\Theta}}_k}^{-\frac{1}{2}}{\boldsymbol
{\tilde \Lambda}}_{{\bf{T}}_k}-\eta_k{\boldsymbol {\tilde
\Lambda}}_{{\boldsymbol{\Theta}}_k}^{-1}\right)^{+}\right]^{\frac{1}{2}}\label{equ:F_opt},
\\
 {\boldsymbol {\Lambda}}_{{\bf{G}_k},{\rm{opt}}}&=
\left[\left(\sqrt{\frac{\gamma_k}{\eta_k\sigma_{n_2}^2}}{\boldsymbol
{\tilde
\Lambda}}_{{\boldsymbol{\Theta}}_k}^{-\frac{1}{2}}{\boldsymbol
{\tilde
\Lambda}}_{{\bf{T}}_k}-\frac{\gamma_k}{\sigma_{n_2}^2}{\boldsymbol
{\tilde
\Lambda}}_{{\boldsymbol{\Theta}}_k}^{-1}\right)^{+}\right]^{\frac{1}{2}}{\boldsymbol
{\tilde \Lambda}}_{{\boldsymbol{\Theta}}_k}^{-\frac{1}{2}},
\label{equ:G_opt}
\end{align} with $\eta_k$ and $\gamma_k$ given by (\ref{eta_k})-(\ref{auxiliary_b}).

From the above summary, it is obvious that the problem of finding
optimal forwarding matrix and equalizer reduces to computing
$P_{r,k}$, and it can be solved based on (\ref{gamma_k}) and the
following two constraints (i.e., (\ref{equ:L_6}) and
(\ref{power_sum}))
\begin{align}
& \gamma_0=\cdots=\gamma_{K-1}, \\
& \sum_k P_{r,k}=P_r.
\end{align}
In the following subsections, we will discuss how to compute
$P_{r,k}$.

\textbf{\textsl{Remark 2:}} When both channels in the two hops are
flat-fading channels, the considered system reduces to
single-carrier AF MIMO relay system. Note that for single-carrier
systems no power allocation has to be calculated since only one
carrier exists, i.e., $P_{r,1}=P_r, K=1$. In this case, the proposed
closed-form solution is exactly the optimal solution for the
transceiver design under channel estimation errors in flat-fading
channel. Furthermore, when the CSI in the two hops are perfectly
known, the derived solution reduces to the optimal solution proposed
in \cite{Guan08}.

 \textbf{\textsl{Remark 3:}} Notice that when the source-relay link
is noiseless and the first hop channel is an identity matrix, the
closed-form solution can be simplified to the optimal linear MMSE
transceiver under channel uncertainties for point-to-point MIMO-OFDM
systems \cite{Rey05}. Moreover, if single carrier transmission is
employed, the closed-form solution further reduces to the optimal
point-to-point MIMO LMMSE transceiver under channel uncertainties
\cite{Ding09}.

 \textbf{\textsl{Remark 4:}} The complexity of the proposed algorithm is dominated
 by one matrix inversion of $(P_{r,k} {\boldsymbol\Psi}
_{rd,k}+\sigma_{n_2}^2{\bf{I}}_{N_R})^{-\frac{1}{2}}$, three matrix
multiplications and one EVD in (\ref{case_2}), one matrix inversion
of ${\bf{R}}_{{\bf{x}}_k}^{-\frac{1}{2}}$, two matrix multiplications and one SVD in (\ref{case_3}), four matrix multiplications in (\ref{F_opt}), four matrix multiplications in (\ref{G_opt}), and two water-filling computations in (\ref{equ:F_opt}) and (\ref{equ:G_opt}). Note that the matrix inversions in (\ref{F_opt}) and (\ref{G_opt}) are the same as those in (\ref{case_2}) and (\ref{case_3}) and therefore their computations could be saved. Specifically, in (\ref{case_2}), the matrix inversion, matrix multiplications and EVD operation have complexities of $O({N_R}^3)$, $O(2{N_R}^3+{N_R}^2{M_D})$ and $O({N_R}^3)$, respectively \cite{Horn85}. In (\ref{case_3}), the matrix inversion, matrix multiplications and SVD operation costs $O({M_R}^3)$, $O({M_R}^2 {N_S}+{M_R}{N_S}^2)$, and $O({M_R}^2 {N_S})$, respectively. With the diagonal structures of ${\boldsymbol {\Lambda}}_{{\bf{F}}_k,{\rm{opt}}}$ and ${\boldsymbol {\Lambda}}_{{\bf{G}_k}, {\rm{opt}}}$, the matrix multiplications in (\ref{F_opt}) and (\ref{G_opt}) have complexities of $O({N_R}^2 {N_k}+{N_R}{N_k}+{N_R}{N_k}{M_R}+{M_R}^2 {N_R})$ and
$O({N_S}{N_k}+{N_S}{N_R}{N_k}+{N_R}^2{N_S}+{N_R}{N_S}{M_D})$, respectively. On the other hand, the complexities for the two water-filling computations in (\ref{equ:F_opt}) and (\ref{equ:G_opt}) are $O(N_k^2)$. As a result, for the AF MIMO-OFDM system with $K$ subcarriers, the complexity of the proposed transceiver design is approximately upper bounded by $O(Km^3)$, where $m={\max\{M_D,N_R,M_R,N_S\}}$.

\subsection{Uncorrelated Channel Estimation Error}

When the channel estimation errors are uncorrelated (for example, by
using training sequences that are white in both time and space
dimensions), the following condition must be satisfied
\cite{Chen08,Ghogho06,Minn06,Stoica03}
\begin{align}
\label{white_training} &{\bf{D}}{\bf{D}}^{\rm{H}} \propto
{\bf{I}}_{N_SL_1}.
\end{align} Then according to (\ref{f_error}), we have
${\boldsymbol \Psi}_{sr,k}= \sum_{\ell_{1}} {\boldsymbol
\Phi}_{\ell_{1},\ell_{1}}^{sr} /K\propto {\bf{I}}_{N_S}$. Similarly,
for the second hop, we also have
\begin{align}
\label{56} {\boldsymbol \Psi}_{rd,k} \propto
{\bf{I}}_{N_R}\triangleq \delta_{rd,k}{\bf{I}}_{N_R},
\end{align} where the specific form of $\delta_{rd,k}$ can be easily derived based on (\ref{Psi_rd}).

Putting (\ref{56}) into the left hand side of (\ref{case_2}), the
expression becomes
\begin{align}
\label{equ_white} &(P_{r,k} {\boldsymbol\Psi}
_{rd,k}+\sigma_{n_2}^2{\bf{I}}_{N_R})^{-\frac{\rm{H}}{2}}{\bf{\hat
H}}_{rd,k}^{\rm{H}}{\bf{\hat H}}_{rd,k}(P_{r,k} {\boldsymbol\Psi}
_{rd,k}+\sigma_{n_2}^2{\bf{I}}_{N_R})^{-\frac{1}{2}}\nonumber
\\
&=\frac{1}{P_{r,k}\delta_{rd,k}+\sigma_{n_2}^2}{\bf{\hat
H}}_{rd,k}^{\rm{H}}{\bf{\hat H}}_{rd,k}.
\end{align}
Applying eigen-decomposition ${\bf{\hat H}}_{rd,k}^{\rm{H}}{\bf{\hat
H}}_{rd,k}= {\bf{U}}_{{\bf{H}}_k}{\boldsymbol
{\Lambda}}_{{\bf{H}}_k}{\bf{U}}_{{\bf{H}}_k}^{\rm{H}}$ and comparing
with the right hand side of (\ref{case_2}), we have
\begin{align}
\label{U_A} {\bf{U}}_{{\boldsymbol
\Theta}_k}&={\bf{U}}_{{\bf{H}}_k},\ \ \ \  {\boldsymbol
{\Lambda}}_{{\boldsymbol{\Theta}}_k}=\frac{1}{(P_{r,k}\delta_{rd,k}+\sigma_{n_2}^2)}{\boldsymbol
{\Lambda}}_{{\bf{H}}_k}.
\end{align}Substituting (\ref{U_A}) into (\ref{gamma_k}), $\gamma_k$ reduces to
\begin{align}
\label{gamma_white}
\gamma_k=\frac{\sigma_{n_2}^2\left({\rm{Tr}}({\boldsymbol {\tilde
\Lambda}}_{{\bf{T}}_k}{\boldsymbol {\tilde
\Lambda}}_{{\bf{H}}_k}^{-\frac{1}{2}}{\boldsymbol
\Lambda}_{{\bf{I}},k})\right)^2}{\left(P_{r,k}\left(1+\delta_{rd,k}{\rm{Tr}}({\boldsymbol
{\tilde \Lambda}}_{{\bf{H}}_k}^{-1}{\boldsymbol
\Lambda}_{{\bf{I}},k})\right)+\sigma_{n_2}^2{\rm{Tr}}({\boldsymbol
{\tilde \Lambda}}_{{\bf{H}}_k}^{-1}{\boldsymbol
\Lambda}_{{\bf{I}},k})\right)^2},
\end{align} where ${\boldsymbol{\tilde
\Lambda}}_{{\bf{H}}_k}$ is the $N_k \times N_k$ principal submatrix
of ${\boldsymbol {\Lambda}}_{{\bf{H}}_k}$.

With (\ref{gamma_white}) and the facts that $\sum_k P_{r,k}=P_r$ and
$\gamma_{0}=\cdots=\gamma_{K-1}$, $P_{r,k}$ can be straightforwardly
computed to be
\begin{align}
\label{P_rk}
&P_{r,k}=\sqrt{\frac{\sigma_{n_2}^2}{\rho}}\frac{{\rm{Tr}}({\boldsymbol
{\tilde \Lambda}}_{{\bf{T}}_k}{\boldsymbol {\tilde
\Lambda}}_{{\bf{H}}_k}^{-\frac{1}{2}}{\boldsymbol
\Lambda}_{{\bf{I}},k})}{{1+ \delta_{rd,k}{\rm{Tr}}({\boldsymbol
{\tilde \Lambda}}_{{\bf{H}}_k}^{-1}{\boldsymbol
\Lambda}_{{\bf{I}},k}) }}-\frac{\sigma_{n_2}^2{\rm{Tr}}({\boldsymbol
{\tilde \Lambda}}_{{\bf{H}}_k}^{-1}{\boldsymbol
\Lambda}_{{\bf{I}},k})}{{1+ \delta_{rd,k}{\rm{Tr}}({\boldsymbol
{\tilde \Lambda}}_{{\bf{H}}_k}^{-1}{\boldsymbol
\Lambda}_{{\bf{I}},k}) }}, \nonumber \\
& k=0, \cdots K-1,
\end{align} where $\rho$ equals to (\ref{solu_gamma}) given at the top of the next page.

\begin{figure*}[!t]
\begin{align}
\label{solu_gamma}& \left. \rho={\sigma_{n_2}^2\left( \sum_k
\frac{{\rm{Tr}}({\boldsymbol {\tilde
\Lambda}}_{{\bf{T}}_k}{\boldsymbol {\tilde
\Lambda}}_{{\bf{H}}_k}^{-\frac{1}{2}}{\boldsymbol
\Lambda}_{{\bf{I}},k})}{1+ \delta_{rd,k}{\rm{Tr}}({\boldsymbol
{\tilde\Lambda}}_{{\bf{H}}_k}^{-1} {\boldsymbol
\Lambda}_{{\bf{I}},k})}
 \right)^2} \middle/ {\left(P_r+\sum_k \frac{\sigma_{n_2}^2{\rm{Tr}}({\boldsymbol
{\tilde \Lambda}}_{{\bf{H}}_k}^{-1}{\boldsymbol
\Lambda}_{{\bf{I}},k})}{1+ \delta_{rd,k}{\rm{Tr}}({\boldsymbol
{\tilde \Lambda}}_{{\bf{H}}_k}^{-1}{\boldsymbol
\Lambda}_{{\bf{I}},k})}\right)^2} \right.
\end{align}
\hrulefill
\end{figure*}

\subsection{Correlated Channel Estimation Error}

Due to limited length of training sequence,
${\bf{D}}{\bf{D}}^{\rm{H}} \propto {\bf{I}}$ may not be possible to
achieve \cite{Stoica03}. In this case, the channel estimation errors
are correlated, and ${\boldsymbol {\Psi}}_{rd,k}\not\propto
{\bf{I}}$. From (\ref{case_2}), it can be seen that the relationship
between ${\boldsymbol \Lambda}_{{\boldsymbol \Theta}_k}$ and
$P_{r,k}$ cannot be expressed in a closed-form . Then the solution
for $P_{r,k}$ cannot be directly obtained. Here, we employ the
spectral approximation (SPA):
\begin{align}
\label{lambda_max}
&P_{r,k} {\boldsymbol\Psi} _{rd,k}+\sigma_{n_2}^2{\bf{I}}_{N_R}
\approx (P_{r,k}\lambda_{\rm{max}}({\boldsymbol
 \Psi}_{rd,k})+\sigma_{n_2}^2){\bf{I}}_{N_R}.
\end{align} For spectral approximation, ${\boldsymbol \Psi}_{rd,k}$ is replaced by $\lambda_{\max}({\boldsymbol \Psi}_{rd,k}){\bf{I}}$, where $\lambda_{\max}({\boldsymbol \Psi}_{rd,k})$ is the maximum eigenvalue of ${\boldsymbol \Psi}_{rd,k}$. Applying (\ref{lambda_max}) to the MSE formulation in (\ref{MSE}), it is obvious that the resultant expression forms an upper-bound to the original MSE. Notice that when the training sequences are close to white sequence
\cite{Andrews2007}, \cite{Ohrman03}, the eigenvalue spread of ${\boldsymbol \Psi}_{rd}$ is small, and SPA is a good approximation. With SPA, the left hand side of (\ref{case_2}) becomes
\begin{align}
\label{equ_nonwhite} &(P_{r,k} {\boldsymbol\Psi}
_{rd,k}+\sigma_{n_2}^2{\bf{I}}_{N_R})^{-\frac{\rm{H}}{2}}{\bf{\hat
H}}_{rd,k}^{\rm{H}}{\bf{\hat H}}_{rd,k}(P_{r,k} {\boldsymbol\Psi}
_{rd,k}+\sigma_{n_2}^2{\bf{I}}_{N_R})^{-\frac{1}{2}} \nonumber \\
 &\approx \frac{1}{P_{r,k}\lambda_{\rm{max}}({\boldsymbol
 \Psi}_{rd,k})+\sigma_{n_2}^2}{\bf{\hat
H}}_{rd,k}^{\rm{H}}{\bf{\hat H}}_{rd,k}.
\end{align}
Comparing (\ref{equ_nonwhite}) to (\ref{equ_white}), it is obvious
that the problem becomes exactly the same as that discussed for
uncorrelated channel estimation errors. Therefore, the allocated
power to the $k^{\rm{th}}$ subcarrier $P_{r,k}$ can be calculated by
(\ref{P_rk}) but with $\delta_{rd,k}$ replaced by
$\lambda_{\max}({\boldsymbol \Psi}_{rd,k})$.

\section{Extension to the Joint Design Involving Source Precoder}
\label{Source_Precoder}
Notice that the design in the previous section is suitable for
scenarios where the source has fixed precoder. For example, the
source precoder can be set to ${\bf{I}}$ for full spatial
multiplexing or space-time block coding matrix for increasing
diversity. On the other hand, if source precoder, relay forwarding
matrix and destination equalizer are jointly designed, we can
proceeds as follows. First, with a source precoder ${\bf{P}}_k$
before transmission, the system model in (\ref{signal_sep}) is
rewritten as
\begin{align}
\label{signal_joint}
{\bf{y}}_k={\bf{H}}_{rd,k}{\bf{F}}_k{\bf{H}}_{sr,k}{\bf{P}}_k{\bf{s}}_k+{\bf{H}}_{rd,k}
{\bf{F}}_k{\bf{n}}_{1,k}+{\bf{n}}_{2,k}.
\end{align} It can be seen that (\ref{signal_joint}) is the same as (\ref{signal_sep}) except ${\bf{H}}_{sr,k}{\bf{P}}_k$ is in the place of ${\bf{H}}_{sr,k}$. Furthermore, without loss of generality, we can assume ${\bf{R}}_{{\bf{s}}_k}={\bf{I}}_{N_k}$ in (\ref{signal_joint}) as all correlations are represented by ${\bf{P}}_k$. Then by using the substitutions ${\bf{H}}_{sr,k} \rightarrow {\bf{H}}_{sr,k}{\bf{P}}_k$ and ${\bf{R}}_{{\bf{s}}_k} \rightarrow {\bf{I}}_{N_k}$ into the first line of (\ref{equ:MSE}), and following the same derivation in Section IV, it can be easily proved that the data MSE at
destination in the $k^{\rm{th}}$ subcarrier is
 \begin{align}
\label{MSE_Joint_A} &{\rm{MSE}}_k({{\bf{G}}}_k,{\bf{F}}_k,{\bf{P}}_k)\nonumber \\
=&\ {\rm{Tr}} \left
( {\bf{G}}_k({\bf{\hat
H}}_{rd,k}{\bf{F}}_k{\bf{R}}_{{\bf{x}}_k}{\bf{F}}_k^{\rm{H}}
{\bf{\hat H}}_{rd,k}^{\rm{H}}+{\bf{K}}_k){\bf{G}}_k^{\rm{H}} \right
)\nonumber \\
& -{\rm{Tr}} \left ( {\bf{P}}_{k}^{\rm{H}}{\bf{\hat
H}}_{sr,k}^{\rm{H}}{\bf{F}}_k^{\rm{H}}{\bf{\hat
H}}_{rd,k}^{\rm{H}}{\bf{G}}_k^{\rm{H}} \right )\nonumber
\\
&-{\rm{Tr}} \left ( {\bf{G}}_k{\bf{\hat
H}}_{rd,k}{\bf{F}}_k{\bf{\hat H}}_{sr,k}{\bf{P}}_k\right
) +{\rm{Tr}} ( {\bf{I}}_{N_k} )
\end{align} where
\begin{align}
{\bf{R}}_{{\bf{x}}_k}&=
{\rm{Tr}}({\bf{P}}_k{\bf{P}}_k^{\rm{H}} {\boldsymbol{\Psi}}
_{sr,k}){\bf{I}}_{M_R} + {\bf{\hat
H}}_{sr,k}{\bf{P}}_k{\bf{P}}_k^{\rm{H}}{\bf{\hat
H}}_{sr,k}^{\rm{H}}+\sigma_{n_1}^2{\bf{I}}_{M_R}.
\end{align}
 Comparing (\ref{MSE}) to (\ref{MSE_Joint_A}), it can be seen that  another way to obtain the data MSE with source precoder is to use the
 substitutions ${\boldsymbol \Psi}_{sr,k}\rightarrow  {\bf{P}}_k^{\rm{H}}{\boldsymbol \Psi}_{sr,k}{\bf{P}}_k$, ${\bf{\hat H}}_{sr,k} \rightarrow {\bf{\hat H}}_{sr,k}{\bf{P}}_k$,
 and ${\bf{R}}_{{\bf{s}}_k} \rightarrow {\bf{I}}_{N_k}$, in
 (\ref{MSE}).

With the additional power constraint for the source precoders, the
optimization problem of joint transceiver design is formulated as
\begin{align}
\label{MSE_Opt_joint}
& \min \limits_{{\bf{G}}_k,{\bf{F}}_k,{\bf{P}}_k}\ \ \sum_k {\rm{MSE}}_k({\bf{G}}_k,{\bf{F}}_k,{\bf{P}}_k) \nonumber \\
& \ \ {\rm{s.t.}} \ \ \ \ \ \ \  \sum_k {\rm{Tr}}({\bf{P}}_{k}{\bf{P}}_{k}^{\rm{H}}) \le P_s \nonumber \\
& \ \ \ \ \ \ \ \ \ \ \ \ \ \sum_k{\rm{Tr}}({\bf{F}}_k{\bf{R}}_{{\bf{x}}_k}{\bf{F}}_k^{\rm{H}})
\le P_r,
\end{align} where $P_s$ is the maximum transmit power at the source.
In general, the optimization problem (\ref{MSE_Opt_joint}) is
nonconvex with respective to the three design variables, and there
is no closed-form solution. However, when ${\bf{P}}_k$'s are fixed,
the solution for ${\bf{G}}_k$'s and ${\bf{F}}_k$'s can be directly
obtained from results given by (\ref{F_opt_aa}) and (\ref{G_opt_aa}) with substitutions ${\boldsymbol \Psi}_{sr,k}\rightarrow
{\bf{P}}_k^{\rm{H}}{\boldsymbol \Psi}_{sr,k}{\bf{P}}_k$, ${\bf{\hat
H}}_{sr,k} \rightarrow {\bf{\hat H}}_{sr,k}{\bf{P}}_k$,
 and ${\bf{R}}_{{\bf{s}}_k} \rightarrow {\bf{I}}_{N_k}$. On the other hand, when ${\bf{G}}_k$'s and
${\bf{F}}_k$'s are fixed, the optimization problem
(\ref{MSE_Opt_joint}) is convex with respect to ${\bf{P}}_k$'s.
Therefore, an iterative algorithm can be employed for joint design
of source precoder, relay forwarding matrix and destination
equalizer.

In order to solve ${\bf{P}}_k$'s when ${\bf{G}}_k$'s and
${\bf{F}}_k$'s are fixed, the data MSE (\ref{MSE_Joint_A}) is
rewritten as
\begin{align}
\label{MSE_k}
&{\rm{MSE}}_k({\bf{G}}_k,{\bf{F}}_k,{\bf{P}}_k)\nonumber \\
=&{\rm{Tr}}\left({\bf{P}}_k^{\rm{H}}({\bf{G}}_k{\bf{\hat
H}}_{rd,k}{\bf{F}}_k{\bf{\hat
H}}_{sr,k})^{\rm{H}}({\bf{G}}_k{\bf{\hat
H}}_{rd,k}{\bf{F}}_k{\bf{\hat H}}_{sr,k}){\bf{P}}_k\right)\nonumber \\
&-{\rm{Tr}}
\left ( {\bf{P}}_{k}^{\rm{H}}{\bf{\hat
H}}_{sr,k}^{\rm{H}}{\bf{F}}_k^{\rm{H}}{\bf{\hat
H}}_{rd,k}^{\rm{H}}{\bf{G}}_k^{\rm{H}} \right )\nonumber
\\
&-{\rm{Tr}} \left ( {\bf{G}}_k{\bf{\hat
H}}_{rd,k}{\bf{F}}_k{\bf{\hat H}}_{sr,k}{\bf{P}}_k\right )+
{\rm{Tr}}({\bf{P}}_k^{\rm{H}}{\bf{N}}_k{\bf{P}}_k)+{\rm{Tr}}({\bf{I}}_{N_k})\nonumber \\
&+a_k,
\end{align} with
\begin{align}
{\bf{N}}_k\triangleq &
{\rm{Tr}}({\bf{G}}_k^{\rm{H}}{\bf{G}}_k)({\bf{\hat
H}}_{sr,k}^{\rm{H}}{\bf{F}}_k^{\rm{H}}{\bf{F}}_k{\bf{\hat
H}}_{sr,k}\lambda_{\max}({\boldsymbol
\Psi}_{rd,k})\nonumber \\
&+\lambda_{\max}({\boldsymbol
\Psi}_{rd,k}){\rm{Tr}}({\bf{F}}_k^{\rm{H}}{\bf{F}}_k){\boldsymbol
\Psi}_{sr,k})\nonumber \\
& +{\rm{Tr}}({\bf{F}}_k^{\rm{H}}{\bf{\hat
H}}_{rd,k}^{\rm{H}}{\bf{G}}_k^{\rm{H}}{\bf{G}}_k{\bf{\hat
H}}_{rd,k}{\bf{F}}_k){\boldsymbol \Psi}_{sr,k}, \label{Psi_rd_approx} \\
a_k=&(\sigma_{n_2}^2+\sigma_{n_1}^2{\rm{Tr}}({\bf{F}}_k{\bf{F}}_k^{\rm{H}}{\boldsymbol
\Psi}_{rd,k})){\rm{Tr}}({\bf{G}}_k{\bf{G}}_k^{\rm{H}})\nonumber \\& +\sigma_{n_1}^2{\rm{Tr}}({\bf{G}}_k{\bf{\hat
H}}_{rd,k}{\bf{F}}_k{\bf{F}}_k^{\rm{H}}{\bf{\hat
H}}_{rd,k}^{\rm{H}}{\bf{G}}_k^{\rm{H}}).
\end{align} In (\ref{Psi_rd_approx}), we have used the spectral
approximation ${\boldsymbol \Psi}_{rd,k} \approx
\lambda_{\max}({\boldsymbol \Psi}_{rd,k}){\bf{I}}_{N_R}$, so that
the objective function for designing ${\bf{P}}_k$'s is consistent
with that of ${\bf{F}}_k$'s and ${\bf{G}}_k$'s. However,
if there is no correlation in the second hop channel estimation
error, ${\boldsymbol \Psi}_{rd,k} = \lambda_{\max}({\boldsymbol
\Psi}_{rd,k} ){\bf{I}}_{N_R}$ and there is no approximation.

Notice that the data MSE (\ref{MSE_k}) is equivalent to the
following expression involving Frobenius norm
\begin{align}
{\rm{MSE}}_k({\bf{G}}_k,{\bf{F}}_k,{\bf{P}}_k)=&\left\|\left[
{\begin{array}{*{20}c}
   {({\bf{G}}_k{\bf{\hat
H}}_{rd,k}{\bf{F}}_k{\bf{\hat H}}_{sr,k}){\bf{P}}_k-{\bf{I}}_{N_k}}  \\
   {{\bf{N}}_k^{1/2}{\bf{P}}_k}  \\
\end{array}}\right] \right\|_F^2\nonumber \\
&+a_k.
\end{align} Furthermore, the two power constraints in the optimization problem (\ref{MSE_Opt_joint}) can
also be reformulated into expressions involving Frobenius norm
\begin{align}
& \|[{\bf{P}}_0^{\rm{T}},\cdots,{\bf{P}}_{K-1}^{\rm{T}}]^{\rm{T}}
\|_F^2 \le P_s. \\
&
\|[({\boldsymbol{\Gamma}}_0{\bf{P}}_0)^{\rm{T}},\cdots,({\boldsymbol{\Gamma}}_{K-1}
{\bf{P}}_{K-1})^{\rm{T}} ]^{\rm{T}}\|_F^2 \le
P_r-\sum_k\sigma_{n_1}^2{\rm{Tr}}({\bf{F}}_k{\bf{F}}_k^{\rm{H}})
\end{align} where
\begin{align}
{\boldsymbol{\Gamma}}_k=({\rm{Tr}}({\bf{F}}_k^{\rm{H}}{\bf{F}}_k)
{\boldsymbol \Psi}_{sr,k}+{\bf{\hat
H}}_{sr,k}^{\rm{H}}{\bf{F}}_k^{\rm{H}}{\bf{F}}_k {\bf{\hat
H}}_{sr,k})^{1/2}.
\end{align}Because the last term $a_k$ in (\ref{MSE_k}) is independent of
${\bf{P}}_k$'s, it can be neglected, and the optimization problem
(\ref{MSE_Opt_joint}) with respective to ${\bf{P}}_k$'s can be
formulated as the following second order conic programming (SOCP)
problem
\begin{align}
& \min_{{\bf{P}}_k,t} \ \ \ t \nonumber \\
& {\rm{s.t.}} \ \nonumber \\
&   \left\| \left[{\begin{array}{*{20}c}
   {({\bf{G}}_0{\bf{\hat
H}}_{rd,0}{\bf{F}}_0{\bf{\hat H}}_{sr,0}){\bf{P}}_0-{\bf{I}}_{N_0}}  \\
   {{\bf{N}}_0^{1/2}{\bf{P}}_0}  \\
  \vdots \\
 {({\bf{G}}_{K-1}{\bf{\hat
H}}_{rd,K-1}{\bf{F}}_{K-1}{\bf{\hat H}}_{sr,K-1}){\bf{P}}_{K-1}-{\bf{I}}_{N_{K-1}}}  \\
   {{\bf{N}}_{K-1}^{1/2}{\bf{P}}_{K-1}}  \\
\end{array}}\right] \right\|_F\le t \nonumber \\
&    \left\| [{\bf{P}}_0^{\rm{T}},\cdots,{\bf{P}}_{K-1}^{\rm{T}}]^{\rm{T}}] \right\|_F\le \sqrt{P_s} \nonumber \\
&
\|[({\boldsymbol{\Gamma}}_0{\bf{P}}_0)^{\rm{T}},\cdots,({\boldsymbol{\Gamma}}_0{\bf{P}}_0)^{\rm{T}}
]^{\rm{T}}\|_{F} \le
\sqrt{P_r-\sum_k\sigma_{n_1}^2{\rm{Tr}}({\bf{F}}_k{\bf{F}}_k^{\rm{H}})}.
\end{align} This problem can be efficiently solved by using inter-point polynomial
algorithms \cite{Boyd04}.

When ${\bf{P}}_k$'s are fixed, the proposed solutions for
${\bf{F}}_k$'s and ${\bf{G}}_k$'s in the previous section are the
optimal solution for the corresponding optimization problem. On the
other hand, when ${\bf{F}}_k$'s and ${\bf{G}}_k$'s are fixed, the
solution for ${\bf{P}}_k$'s obtained from the SOCP problem is
also the optimal solution. It means that the objective
function of joint transceiver design monotonically decreases at each
iteration, and the proposed iterative algorithm converges.

\section{Simulation Results and Discussions}
\label{simulation}

In this section, we investigate the performance of the proposed
algorithms. For the purpose of comparison, the algorithm based on
estimated channel only (without taking the estimation errors into
account) is also simulated. An AF MIMO-OFDM relay system where the
source, relay and destination are equipped with same number of
antennas, $N_S=M_R=N_R=M_D=4$ is considered. The number of
subcarriers $K$ is set to be 64, and the length of the multi-path
channels in both hops is $L=4$. The channel impulse response is generated
according to the HIPERLAN/2 standard \cite{Chen08}. The
signal-to-noise ratio (${\rm{SNR}}$) of the first hop is defined as
${\rm{E}}_s/{\rm{N}}_1=P_s/(K\sigma_{n_1}^2)$, and is fixed as
$30{\rm{dB}}$. At the source, on each subcarrier, four independent
data streams are transmitted, and QPSK is used as the modulation
scheme. The ${\rm{SNR}}$ at the second hop is defined as
${\rm{E}}_r/{\rm{N}}_2=P_r/(K\sigma_{n_2}^2)$. In the figures, MSE
is referred to total simulated MSE over all subcarriers normalized
by $K$.  Each point in the following figures is an average of 10000
realizations. In order to solve SOCP problems, the widely used
optimization matlab toolbox CVX is adopted \cite{Grant07}.

Based on the definition of ${\bf{D}}$ in (\ref{D_Matrix}),
${\bf{D}}{\bf{D}}^{\rm{H}}$ is a block circular matrix. In the
following, only the effect of spatial correlation in training
sequence is demonstrated, and the training is white in time
domain. In this case, ${\bf{D}}{\bf{D}}^{\rm{H}}$ is a block
diagonal matrix, and can be written as  $
{\bf{D}}{\bf{D}}^{\rm{H}}={\bf{I}}_{L} \otimes
\sum_i{\bf{d}}_i{\bf{d}}_i^{\rm{H}}$, where
$\sum_i{\bf{d}}_i{\bf{d}}_i^{\rm{H}}/K$ is the spatial correlation
matrix of the training sequence. Furthermore, the widely used
exponential correlation model is adopted to denote the \begin{figure}[!ht]
\centering
\includegraphics[width=.45\textwidth]{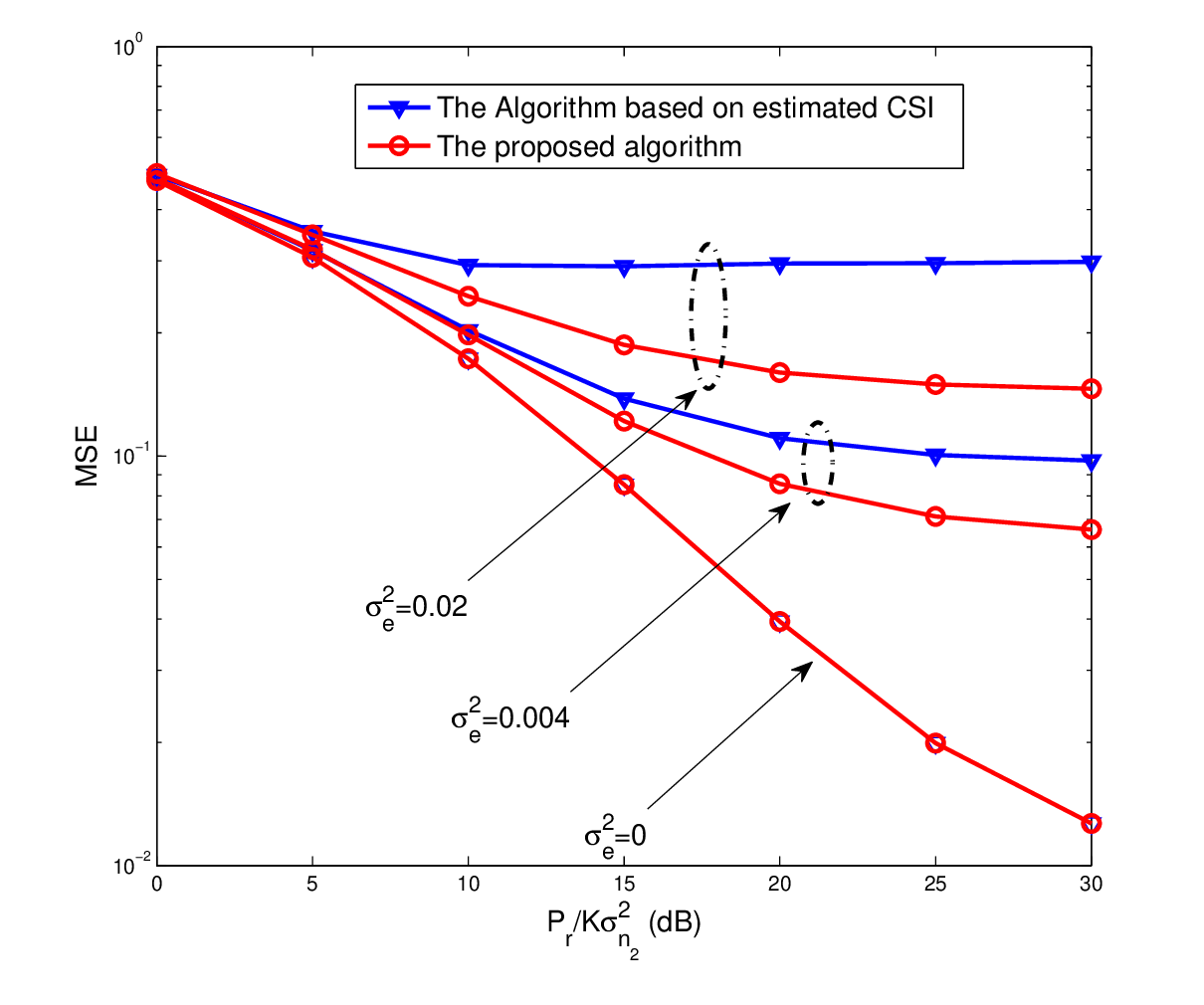}
\caption{MSE of received signal at the destination for different
$\sigma_e^2$ when $\alpha=0.4$ and with ${\bf{P}}_k={\bf{I}}_4$. }\label{fig:2}
\end{figure}spatial
correlation \cite{Ding09}, \cite{Zhang08}, and therefore we
have
\begin{align}
\label{DD} {\bf{D}}{\bf{D}}^{\rm{H}}={\bf{I}}_{L}\otimes K \left[
{\begin{array}{*{20}c}
   {1} & {\alpha} & {\alpha^2} & {\alpha^3}   \\
   {\alpha} & {1} & {\alpha} & {\alpha^2}   \\
     {\alpha^2}& {\alpha} & {1} & {\alpha}  \\
     {\alpha^3} &  {\alpha^2}& {\alpha} & {1} \\
\end{array}} \right].
\end{align} It is assumed that the same training
sequence is used for channel estimation in the two hops. Based on
the definition of ${\boldsymbol \Psi}_{sr,k}$ and ${\boldsymbol
\Psi}_{rd,k}$ in (\ref{Psi_sr}) and (\ref{Psi_rd}), and together
with (\ref{DD}), we have
\begin{align}
\label{simu_a} {\boldsymbol \Psi}_{sr,k}={\boldsymbol
\Psi}_{rd,k}=\sigma_e^2 \left[ {\begin{array}{*{20}c}
   {1} & {\alpha} & {\alpha^2} & {\alpha^3}   \\
   {\alpha} & {1} & {\alpha} & {\alpha^2}   \\
     {\alpha^2}& {\alpha} & {1} & {\alpha}  \\
     {\alpha^3} &  {\alpha^2}& {\alpha} & {1} \\
\end{array}} \right]^{-1},
\end{align}  where $\sigma_e^2=1/{\rm{SNR}}_e$ can be viewed as the variance of
channel estimation errors and ${\rm{SNR}}_e$ is SNR during channel estimation process.

First, we investigate the performance of the proposed algorithm with fixed source precoder ${\bf{P}}_k={\bf{I}}_4$ and when
$\alpha=0.4$ in (\ref{simu_a}). Fig.~\ref{fig:2} shows the MSE of
the received signal at the destination with different $\sigma_e^2$.
It can be seen that the performance of the proposed algorithm is
always better than that of the algorithm based on estimated CSI
only, as long as $\sigma_e^2$ is not zero. Furthermore, the
performance improvement of the proposed algorithm over the algorithm
based on only estimated CSI enlarges when $\sigma_e^2$ increases.

\begin{figure}[!ht]
\centering
\includegraphics[width=.45\textwidth]{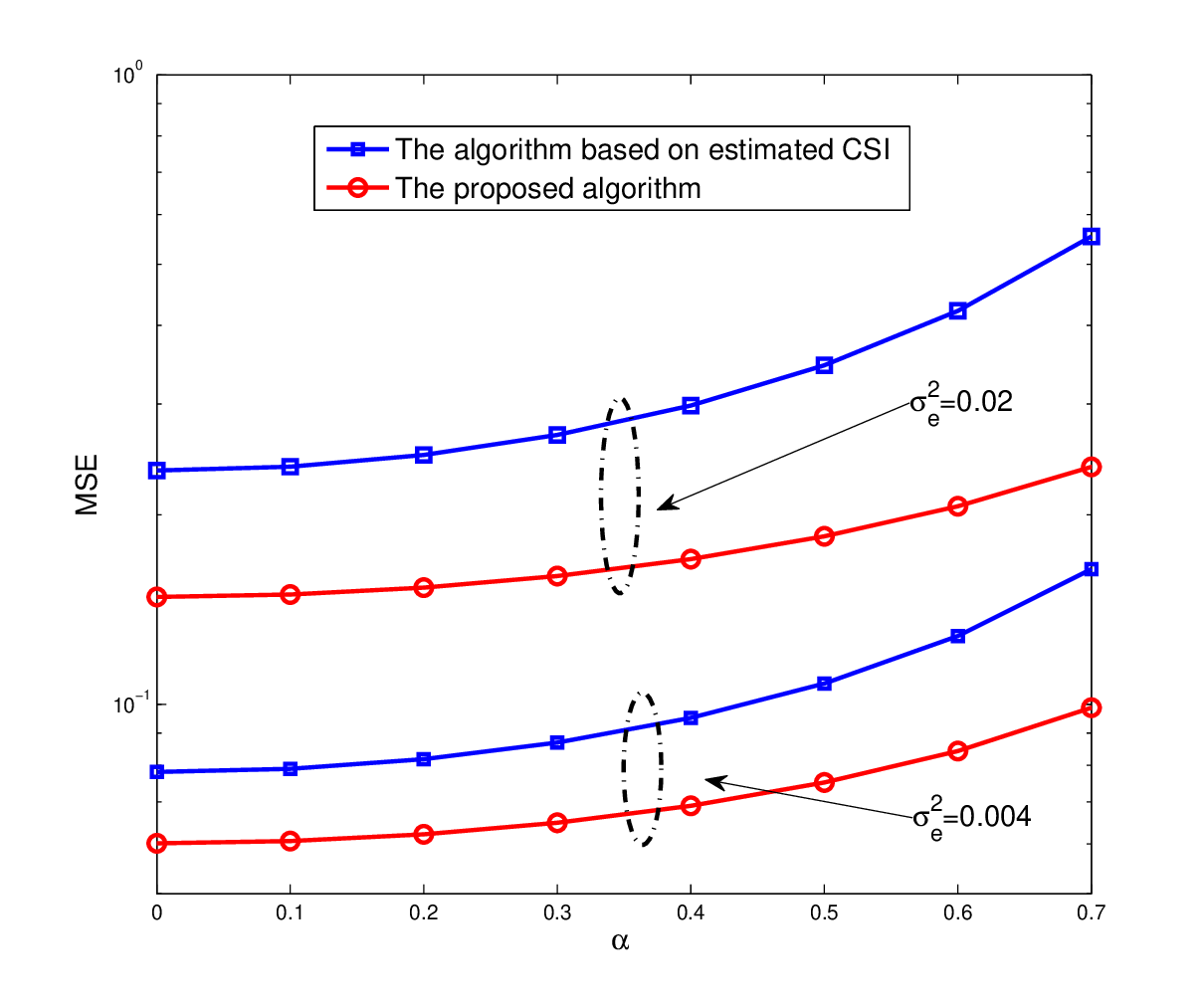}
\caption{MSE of received signal at the destination for different
$\alpha$ when $P_r/K\sigma_{n_2}^2$=30dB and with ${\bf{P}}_k={\bf{I}}_4$. }\label{fig:3}
\end{figure}

Fig.~\ref{fig:3} shows the MSE of the output
data at the destination
for both proposed algorithm and the algorithm based on estimated CSI
only with fixed source precoder ${\bf{P}}_k={\bf{I}}_4$ and with different $\alpha$. It can be seen that although
performance degradation is observed for both algorithms when
$\alpha$ increases, the proposed algorithm shows a significant
improvement over the algorithm based on estimated CSI only. Furthermore, as $\alpha=0$ gives the best data MSE
performance, it demonstrates that white sequence is preferred in
channel estimation.

\begin{figure}[!ht]
\centering
\includegraphics[width=.45\textwidth]{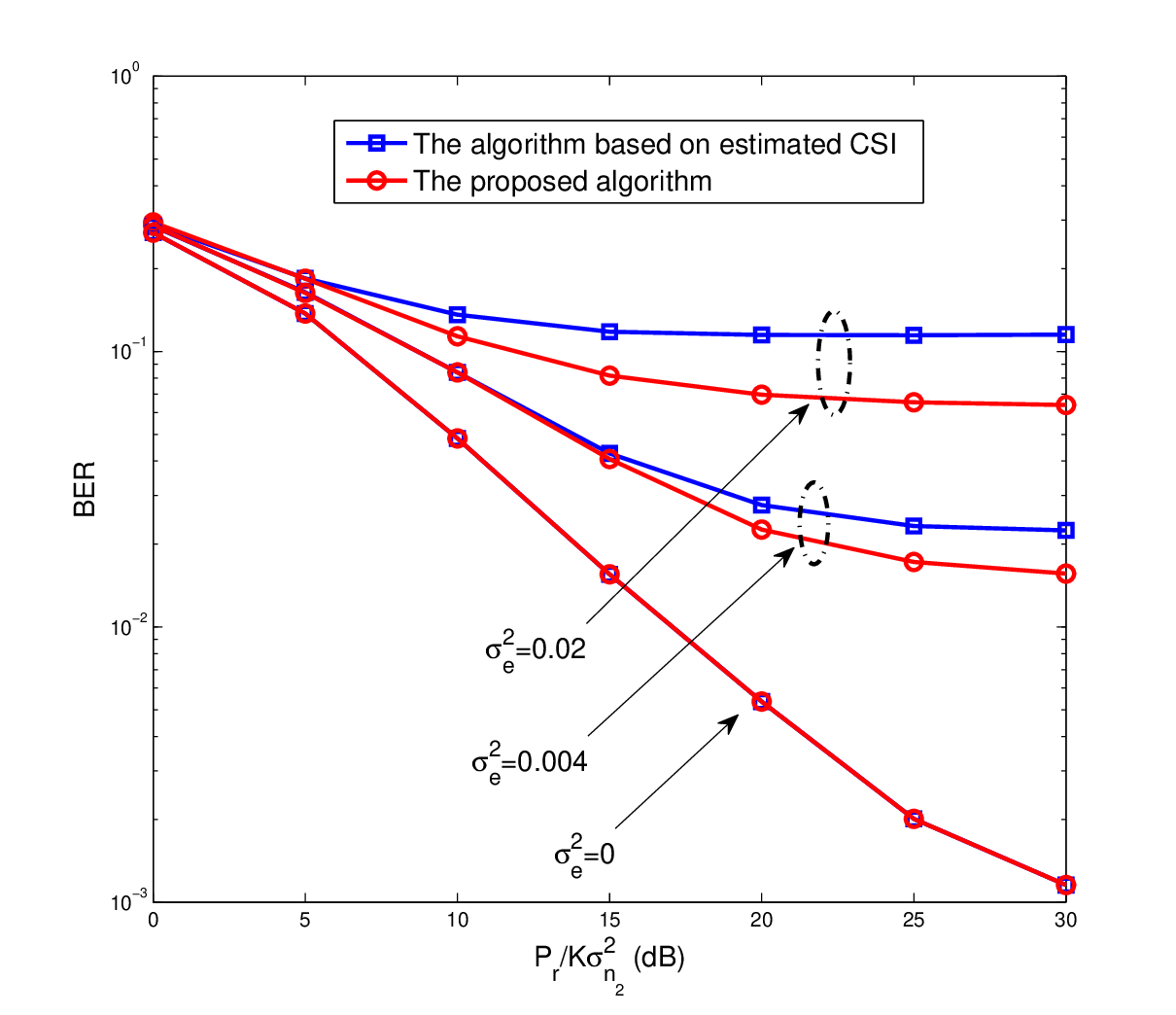}
\caption{BER of received data at the destination for different
$\sigma_e^2$ when $\alpha=0.5$ and with ${\bf{P}}_k={\bf{I}}_4$. }\label{fig:4}
\end{figure}

Fig.~\ref{fig:4} shows the bit error rates (BER) of the
output data at the destination for different $\sigma_e^2$, when
$\alpha=0.5$.  \begin{figure}[!ht]
\centering
\includegraphics[width=.45\textwidth]{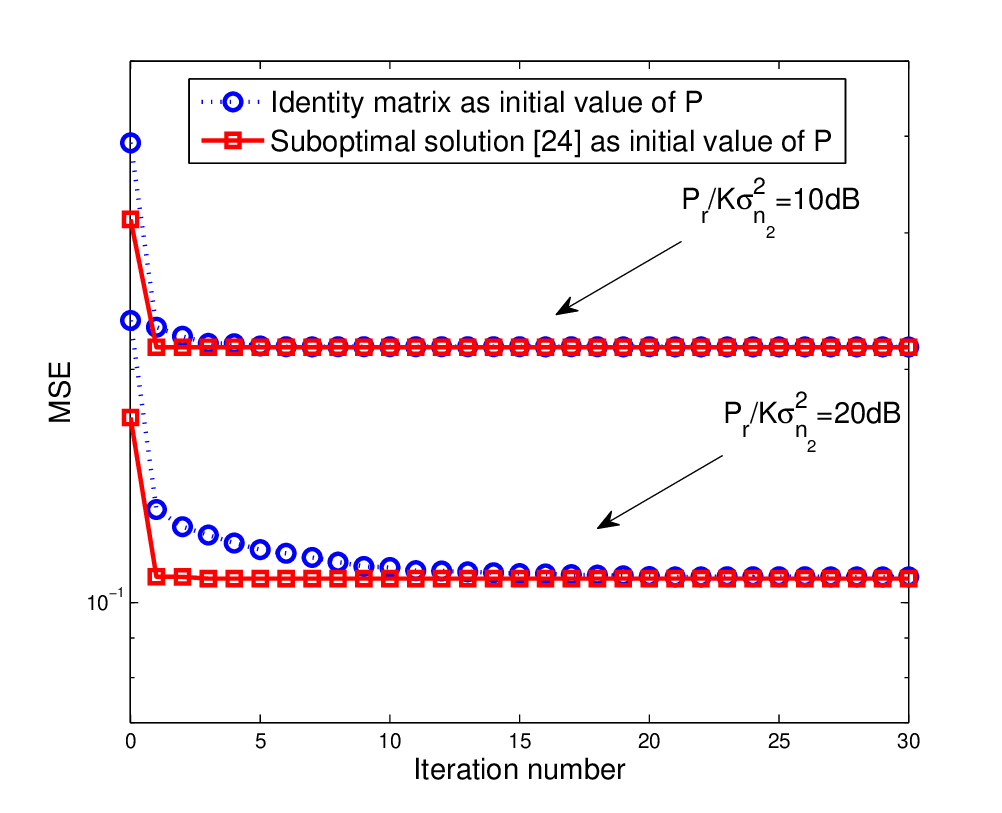}
\caption{Convergence behavior of the proposed iterative algorithm when
$\alpha=0.4$ and $\sigma_e^2=0.01$.}\label{fig:5}
\end{figure}It can be seen that the BER performance is consistent
with MSE performance in Fig.~\ref{fig:2}.

When source precoder design is considered, the proposed algorithm is an iterative algorithm.
Fig.~\ref{fig:5} shows the convergence behavior of the proposed iterative algorithm with different initial values
of ${\bf{P}}$. In the figure, the suboptimal solution as the initial
value for ${\bf{P}}$ refers to the solution given in
\cite{Rey05} based on the first hop CSI. It can be seen that the
proposed algorithm with suboptimal solution as initial value has a
faster convergence speed than that with identity matrix as the
initial value.

\begin{figure}[!ht]
\centering
\includegraphics[width=.45\textwidth]{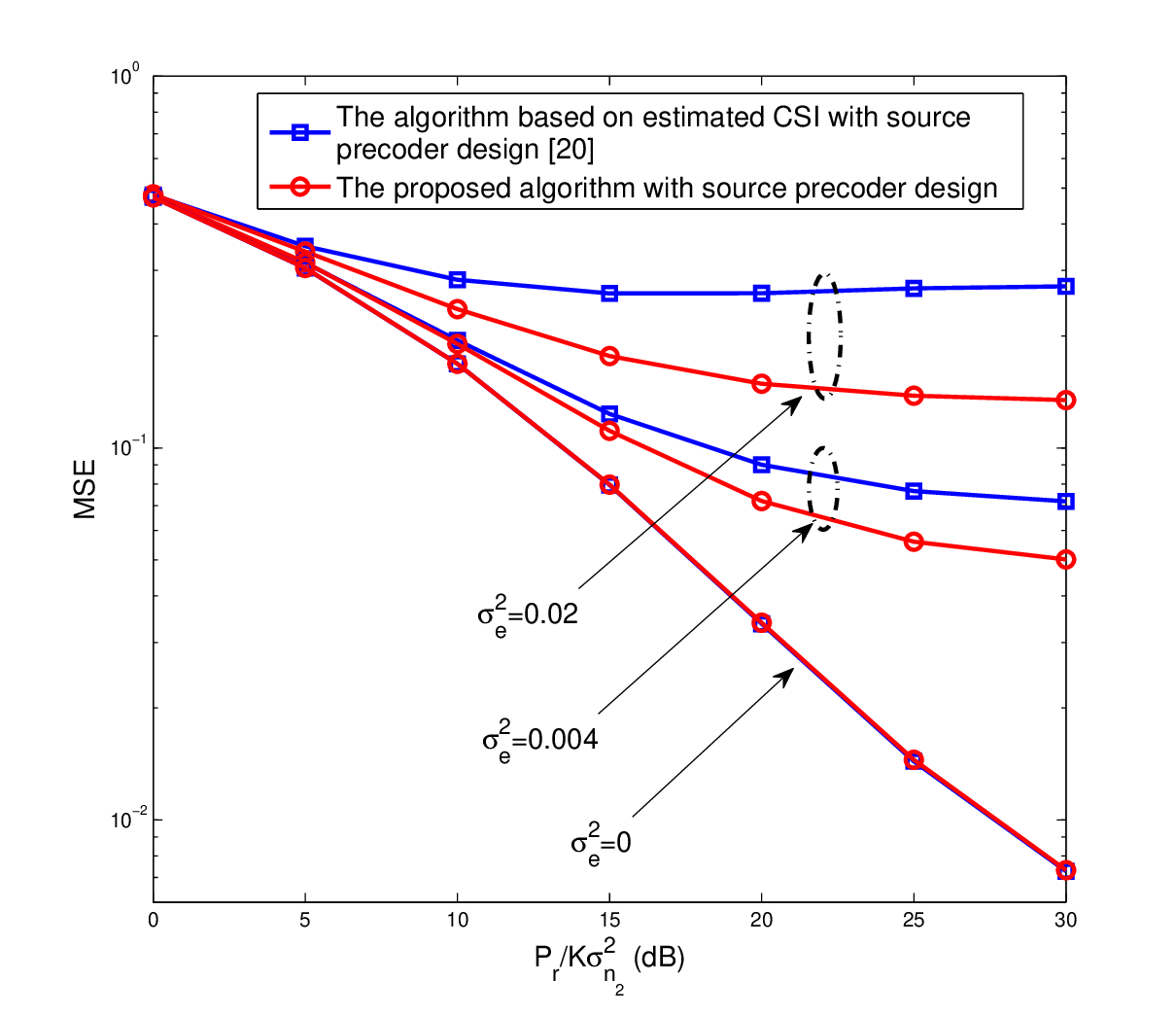}
\caption{MSE of received signal at the destination for different
$\sigma_e^2$ when $\alpha=0.4$. }\label{fig:6}
\end{figure}

Fig.~\ref{fig:6} compares the data MSEs of the proposed iterative
algorithm under channel uncertainties and the iterative algorithm based on
estimated CSI only in \cite{Rong09}. Similar to the case with fixed source precoder, the proposed joint design algorithm taking into account the channel estimation uncertainties performs better than the algorithm based on estimated CSI only.

\begin{figure}[!ht]
\centering
\includegraphics[width=.45\textwidth]{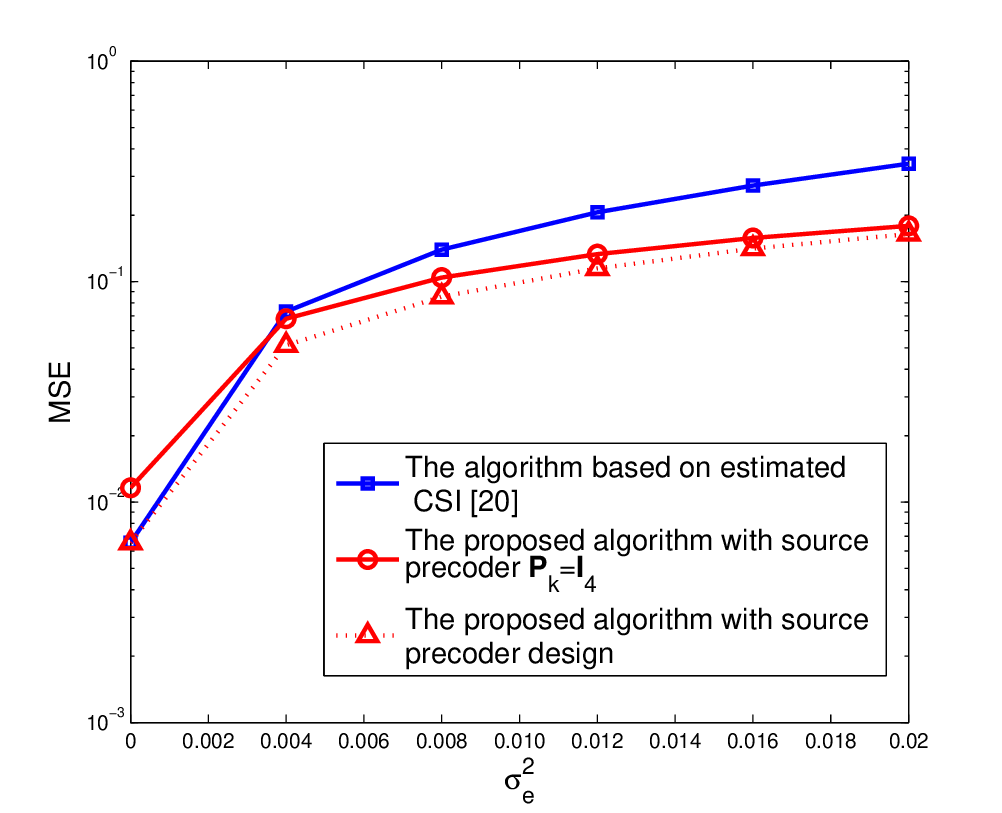}
\caption{MSE of received data at the destination for different
$\sigma_e^2$, when $\alpha=0.5$ and
$P_r/K\sigma_{n_2}^2$=30dB.}\label{fig:7}
\end{figure}
Finally, Fig.~\ref{fig:7} illustrates the data MSE of the iterative transceiver design algorithm
based on estimated CSI only \cite{Rong09} and the proposed algorithms with source precoder jointly designed or simply set to ${\bf{P}}_k={\bf{I}}_4$. It can be seen that when CSI is
perfectly known ($\sigma_e^2=0$), the algorithms with source precoder design performs better than that by setting precoder ${\bf{P}}_k={\bf{I}}_4$. On the other hand, when
$\sigma_e^2\ge 0.004$, even the proposed algorithm with simple precoder ${\bf{P}}_k={\bf{I}}_4$ performs better than the algorithm based on estimated CSI only with source precoder design. Furthermore, when the channel
estimation errors increases, the performance gap between
the proposed algorithms with and without source precoder design decreases.  Notice that the algorithm without source precoder design has a much lower complexity, thus it represents a promising tradeoff in terms of complexity and performance.

\section{Conclusions}
\label{conclusions} In this paper, linear transceiver design was addressed for AF MIMO-OFDM
relaying systems with channel estimation errors based on MMSE
criterion. The linear channel estimators and the corresponding MSE
expressions were first derived. Then a general solution for optimal
relay forwarding matrix and destination equalizer was proposed. When the channel estimation
errors are uncorrelated, the optimal solution is in closed-form, and
it includes several existing transceiver design results as special
cases. Furthermore, the design was extended to the case where source precoder design is involved. Simulation results
showed that the proposed algorithms offer significant performance
improvements over the algorithms based on estimated CSI only.

\appendices

\section{Proof of (\ref{signalmodel})}
\label{app_sigmal_model}

Based on the characteristics of DFT operation, the matrix
${\boldsymbol{\mathcal{H}}}_{sr}$ defined in (\ref{signal_time}) is
a $KM_R \times KN_S $ block circulant matrix given by (\ref{a_equ_1}) at the top of the next page,
\begin{figure*}[!t]
\begin{align}
\label{a_equ_1}
{\boldsymbol{\mathcal{H}}}_{sr} \triangleq \left[
{\begin{array}{*{20}c}
   {{\boldsymbol{\mathcal{H}}}_{sr}^{(0)}} & {\bf{0}} & {\bf{0}} &   \cdots  & {{\boldsymbol{\mathcal{H}}}_{sr}^{(L_1-1)}} & {{\boldsymbol{\mathcal{H}}}_{sr}^{(L_1-2)}}  &  \cdots  & {{\boldsymbol{\mathcal{H}}}_{sr}^{(1)}} \\
   {{\boldsymbol{\mathcal{H}}}_{sr}^{(1)}} &  {{\boldsymbol{\mathcal{H}}}_{sr}^{(0)}}  & {\bf{0}}  &  \cdots & {\bf{0}}& {{\boldsymbol{\mathcal{H}}}_{sr}^{(L_1-1)}} &  \cdots  & {{\boldsymbol{\mathcal{H}}}_{sr}^{(2)}}\\
    \vdots  &  \vdots  &  \vdots  &  \cdots  &  \vdots  &  \vdots &  \vdots  &  \vdots   \\
    {\bf{0}}&  \cdots  & {\bf{0}}  &{{\boldsymbol{\mathcal{H}}}_{sr}^{(L_1-1)}}  & {{\boldsymbol{\mathcal{H}}}_{sr}^{(L_1-2)}}&{{\boldsymbol{\mathcal{H}}}_{sr}^{(L_1-3)}} &  \cdots  & {{\boldsymbol{\mathcal{H}}}_{sr}^{(0)}}  \\
\end{array}} \right]
\end{align}
\hrulefill
\end{figure*}whose element
${\boldsymbol{\mathcal{H}}}_{sr}^{(\ell)}$ is defined in
(\ref{H_sr_time}). It is obvious that
${\boldsymbol{\mathcal{H}}}_{sr}^{(\ell)}$ is the $\ell^{\rm{th}}$
tap of the multi-path MIMO channels between the source and relay in
the time domain and $L_1$ is the length of the multi-path channel.

On the other hand, based on the definition of ${\bf{d}}$ in
(\ref{signal_time}), we have the relationship between  ${\bf{d}}$ and  ${\bf{s}}$ which is given by (\ref{a_equ_2}).
\begin{figure*}[!t]
\begin{align}
\label{a_equ_2}
{\bf{d}}&=[(\underbrace{\frac{1}{\sqrt{K}}\sum_{k=0}^{K-1}{\bf{s}}_{k}e^{j\frac{2\pi}{K}k(0)}}_{\bf{d}_0})^{\rm{T}}
\
(\underbrace{\frac{1}{\sqrt{K}}\sum_{k=0}^{K-1}{\bf{s}}_{k}e^{j\frac{2\pi}{K}k(1)}}_{\bf{d}_1})^{\rm{T}}\
\cdots \
(\underbrace{\frac{1}{\sqrt{K}}\sum_{k=0}^{K-1}{\bf{s}}_{k}e^{j\frac{2\pi}{K}k(K-1)}}_{{\bf{d}}_{K-1}})^{\rm{T}}]^{\rm{T}}
\end{align}\hrulefill
\end{figure*} From (\ref{a_equ_1}) and (\ref{a_equ_2}), by straightforward computation, the signal
model given in (\ref{signal_time}) can be reformulated as
\begin{align}
{\bf{r}}&={\boldsymbol{\mathcal{H}}}_{sr}{\bf{d}}+{\bf{v}}\nonumber \\
&={{\rm{vec}}}([ {{\boldsymbol{\mathcal{H}}}_{sr}^{(0)}} \   \cdots
\ {{\boldsymbol{\mathcal{H}}}_{sr}^{(L_1-1)}} ]{\bf{D}})+{\bf{v}}
\nonumber \\
&=({\bf{D}}^{\rm{T}} \otimes {\bf{I}}_{M_R}) {{\rm{vec}}}([
{{\boldsymbol{\mathcal{H}}}_{sr}^{(0)}} \ \cdots \
{{\boldsymbol{\mathcal{H}}}_{sr}^{(L_1-1)}} ])+{\bf{v}},
\end{align} where the matrix ${\bf{D}}$ is defined in (\ref{D_Matrix}).

\section{Proof of (\ref{Expectaion_H_sr})}
\label{Lemma}

For the expectation of the following product
\begin{align}
{\boldsymbol{\Sigma}}={\mathbb{E}}\{
{\bf{Q}}{\bf{R}}{\bf{W}}^{\rm{H}} \}
\end{align} where ${\bf{Q}}$ and ${\bf{W}}$
are two $M \times N $ random matrices with compatible dimension to
${\bf{R}}$, the $(i,j)^{\rm{th}}$ element of ${\boldsymbol{\Sigma}}$
is
\begin{align}
\label{sigma}
{\boldsymbol{\Sigma}}({i,j})&={\mathbb{E}}\{{\bf{Q}}({i,:}){\bf{R}}{\bf{W}}({j,:})^{\rm{H}}\}
\nonumber \\
&=\sum_{t}\sum_{k}{\mathbb{E}}\{{\bf{Q}}({i,t}){\bf{R}}({t,k}){\bf{W}}({j,k})^*\}.
\end{align}
If the two random matrices ${\bf{Q}}$ and ${\bf{W}}$ satisfy
\begin{align}
&{\mathbb{E}}\{ {\rm{vec}}({\bf{Q}}){\rm{vec}}^{\rm{H}}({\bf{W}})
\}={\bf{A}}\otimes {\bf{B}},
\end{align} where ${\bf{A}}$ is a $N \times N$ matrix while
${\bf{B}}$ is a $M \times M$ matrix, then we have the
equality $
{\mathbb{E}}\{{\bf{Q}}({i_1,j_1}){\bf{W}}({i_2,j_2})^{*}\}={\bf{B}}({i_1,i_2}){\bf{A}}({j_1,j_2})$.
As ${\bf{Q}}({i,t})$ and ${\bf{W}}({j,k})$ are scalars,
(\ref{sigma}) can be further written as
\begin{align}
\label{92}
{\boldsymbol{\Sigma}}({i,j})=& \sum_{t}\sum_{k}({\bf{R}}({t,k}){\mathbb{E}}\{{\bf{Q}}({i,t}){\bf{W}}({j,k})^*\}) \nonumber \\
=& \sum_{t}\sum_{k}{\bf{R}}({t,k}){\bf{A}}({t,k}){\bf{B}}({i,j}).
\end{align}Finally,
writing (\ref{92}) back to matrix form, we have \cite{Kay93}
\begin{align}
\label{Expectation}
{\boldsymbol{\Sigma}}&={\bf{B}}{\rm{Tr}}({\bf{R}}{\bf{A}}^{\rm{T}}).
\end{align} Notice that this conclusion is independent of the matrix
variate distributions of  ${\bf{Q}}$ and ${\bf{W}}$, but only
determined by their second order moments. Putting
${\bf{A}}=\sum_{\ell_2=0}^{L_1-1}\sum_{\ell_1=0}^{L_1-1}(e^{-j\frac{2\pi}{K}k(\ell_{1}-\ell_{2})}{\boldsymbol{\Phi}}_{\ell_{1},\ell_{2}}^{sr})$,
${\bf{B}}={\bf{I}}_{M_R}$ and
${\bf{Q}}={\bf{W}}=\Delta{\bf{H}}_{sr,k}$ , into
(\ref{Expectation}), we have (\ref{Expectaion_H_sr}).

\section{Proof of Property 1}
\label{Lemma_1}

Right multiplying both sides of (\ref{equ:L_1}) with
${\bf{G}}_k^{\rm{H}}$, the following equality holds
\begin{align}
\label{38} &{\bf{G}}_k({\bf{\hat
H}}_{rd,k}{\bf{F}}_k{\bf{R}}_{{\bf{x}}_k}{\bf{F}}_k^{\rm{H}}
{\bf{\hat H}}_{rd,k}^{\rm{H}}+{\bf{K}}_k){\bf{G}}_k^{\rm{H}}\nonumber \\
&=
{\bf{R}}_{s_k}({\bf{\hat H}}_{rd,k}{\bf{F}}_k{\bf{\hat
H}}_{sr,k})^{\rm{H}}{\bf{G}}_k^{\rm{H}}.
\end{align}
Left multiplying (\ref{equ:L_2}) with ${\bf{F}}_k^{\rm{H}}$, we have
\begin{align}
\label{39} & {\bf{F}}_k^{\rm{H}}{\bf{\hat
H}}_{rd,k}^{\rm{H}}{\bf{G}}_k^{\rm{H}}{\bf{G}}_k{\bf{\hat
H}}_{rd,k}{\bf{F}}_k{\bf{R}}_{{\bf{x}}_k}+
{\bf{F}}_k^{\rm{H}}{\rm{Tr}}({\bf{G}}_k{\bf{G}}_k^{\rm{H}}){\boldsymbol\Psi}
_{rd,k}{\bf{F}}_k{\bf{R}}_{{\bf{x}}_k}\nonumber \\
& +\gamma_k{\bf{F}}_k^{\rm{H}}{\bf{F}}_k{\bf{R}}_{{\bf{x}}_k} = {\bf{F}}_k^{\rm{H}}({\bf{\hat
H}}_{sr,k}{\bf{R}}_{s_k}{\bf{G}}_k{\bf{\hat
H}}_{rd,k})^{\rm{H}}.
\end{align}
After taking the traces of both sides of (\ref{38}) and (\ref{39})
and with the fact that the traces of their righthand sides are
equivalent, i.e.,
\[
{\rm{Tr}}({\bf{R}}_{s_k}({\bf{\hat
H}}_{rd,k}{\bf{F}}_k{\bf{\hat
H}}_{sr,k})^{\rm{H}}{\bf{G}}_k^{\rm{H}})={\rm{Tr}}({\bf{F}}_k^{\rm{H}}({\bf{\hat
H}}_{sr,k}{\bf{R}}_{s_k}{\bf{G}}_k{\bf{\hat H}}_{rd,k})^{\rm{H}}),\]
we directly have
\begin{align}
\label{equal} &{\rm{Tr}}({\bf{G}}_k({\bf{\hat
H}}_{rd,k}{\bf{F}}_k{\bf{R}}_{{\bf{x}}_k}{\bf{F}}_k^{\rm{H}}
{\bf{\hat
H}}_{rd,k}^{\rm{H}}+{\bf{K}}_k){\bf{G}}_k^{\rm{H}}) \nonumber \\
&={\rm{Tr}}({\bf{F}}_k^{\rm{H}}{\bf{\hat
H}}_{rd,k}^{\rm{H}}{\bf{G}}_k^{\rm{H}}{\bf{G}}_k{\bf{\hat
H}}_{rd,k}{\bf{F}}_k{\bf{R}}_{{\bf{x}}_k})\nonumber \\
& +\gamma_k{\rm{Tr}}({\bf{F}}_k^{\rm{H}}{\bf{F}}_k{\bf{R}}_{{\bf{x}}_k})
+ {\rm{Tr}}({\bf{G}}_k{\bf{G}}_k^{\rm{H}})
{\rm{Tr}}({\bf{F}}_k^{\rm{H}}{\boldsymbol\Psi}
_{rd,k}{\bf{F}}_k{\bf{R}}_{{\bf{x}}_k} ).
\end{align} By the property of trace operator,
 \begin{align}
 &{\rm{Tr}}({\bf{G}}_k({\bf{\hat
H}}_{rd,k}{\bf{F}}_k{\bf{R}}_{{\bf{x}}_k}{\bf{F}}_k^{\rm{H}}
{\bf{\hat
H}}_{rd,k}^{\rm{H}}){\bf{G}}_k^{\rm{H}})\nonumber \\
&={\rm{Tr}}({\bf{F}}_k^{\rm{H}}{\bf{\hat
H}}_{rd,k}^{\rm{H}}{\bf{G}}_k^{\rm{H}}{\bf{G}}_k{\bf{\hat
H}}_{rd,k}{\bf{F}}_k{\bf{R}}_{{\bf{x}}_k}), \nonumber
\end{align}
 and (\ref{equal})
reduces to
\begin{align}
\label{Comp_1} &{\rm{Tr}}({\bf{G}}_k{\bf{K}}_k{\bf{G}}_k^{\rm{H}})=
{\rm{Tr}}({\bf{G}}_k{\bf{G}}_k^{\rm{H}})
{\rm{Tr}}({\bf{F}}_k^{\rm{H}}{\boldsymbol\Psi}
_{rd,k}{\bf{F}}_k{\bf{R}}_{{\bf{x}}_k} ) \nonumber \\
& +\gamma_k{\rm{Tr}}({\bf{F}}_k^{\rm{H}}{\bf{F}}_k{\bf{R}}_{{\bf{x}}_k}).
\end{align}

On the other hand, based on the definition of ${\bf{K}}_k$ in
(\ref{K}), ${\rm{Tr}}({\bf{G}}_k{\bf{K}}_k{\bf{G}}_k^{\rm{H}})$ can
be also expressed as
\begin{align}
\label{Comp_2} &{\rm{Tr}}({\bf{G}}_k{\bf{K}}_k{\bf{G}}_k^{\rm{H}})=
{\rm{Tr}}({\bf{G}}_k{\bf{G}}_k^{\rm{H}})
{\rm{Tr}}({\bf{F}}_k^{\rm{H}}{\boldsymbol\Psi}
_{rd,k}{\bf{F}}_k{\bf{R}}_{{\bf{x}}_k} )\nonumber
\\&
+{\rm{Tr}}({\bf{G}}_k{\bf{R}}_{n_2,k}{\bf{G}}_k^{\rm{H}}).
\end{align} Comparing (\ref{Comp_1}) with (\ref{Comp_2}), it can be
concluded that
\begin{align}
\label{109} {\rm{Tr}}({\bf{G}}_k{\bf{R}}_{n_2,k}{\bf{G}}_k^{\rm{H}})
=\gamma_k{\rm{Tr}}({\bf{F}}_k{\bf{R}}_{{\bf{x}}_k}{\bf{F}}_k^{\rm{H}}).
\end{align} Putting (\ref{109}) into (\ref{equ:L_3}), we have
${\rm{Tr}}({\bf{G}}_k{\bf{R}}_{n_2,k}{\bf{G}}_k^{\rm{H}})-\gamma_k
P_{r,k}=0$. As
${\bf{R}}_{{{n}}_{2,k}}=\sigma_{n_2}^2{\bf{I}}_{M_D}$, it is
straightforward that
\begin{align}
\label{gamma}
\sigma_{n_2}^2{{\rm{Tr}}({\bf{G}}_k{\bf{G}}_k^{\rm{H}})}=\gamma_k{P_{r,k}}.
\end{align}

Furthermore, based on the fact $ \gamma_0=\gamma_1=\cdots=\gamma_{K-1}=\rho$
and taking summation of both sides of (\ref{gamma}), the following
equation holds
\begin{align}
\label{100} &\sum_k
\sigma_{n_2}^2{\rm{Tr}}({\bf{G}}_k{\bf{G}}_k^{\rm{H}})=\rho \sum_k
P_{r,k}.
\end{align}
Putting (\ref{100}) into (\ref{equ:L_5}), we have
\begin{align}
\sum_k\sigma_{n_2}^2{\rm{Tr}}({\bf{G}}_k{\bf{G}}_k^{\rm{H}})-\rho
P_r=0,
\end{align} and it follows that
\begin{align}
\gamma_k=\rho=\sigma_{n_2}^2\frac{\sum_k{\rm{Tr}}({\bf{G}}_k{\bf{G}}_k^{\rm{H}})}{P_r}.
\end{align}

Since for the optimal equalizer ${\bf{G}}_k$,
$\sum_{k}{\rm{Tr}}({\bf{G}}_{k,{\rm{opt}}}{\bf{G}}_{k,{\rm{opt}}}^{\rm{H}})
\ne 0$, it can be concluded that $\gamma_k \ne 0$. In order to have
(\ref{equ:L_3}) satisfied, we must have
\begin{align}
\label{equ_FF}
{\rm{Tr}}({\bf{F}}_{k,{\rm{opt}}}{\bf{R}}_{{\bf{x}}_k}{\bf{F}}_{k,{\rm{opt}}}^{\rm{H}})=P_{r,k}.
\end{align} Furthermore, as $\rho \ne 0$, based on (\ref{equ:L_5}), it is also concluded that
\begin{align}
\sum_k P_{r,k}=P_r.
\end{align}
Finally, (\ref{gamma}) constitutes the second part of the Property 1.

\section{Proof of Property 2}
\label{Lemma2}

Defining a full rank Hermitian matrix $
{\bf{M}}_{k}=P_{r,k}{\boldsymbol \Psi}_{rd,k}+{\sigma}_{n_2}^2
{\bf{I}}_{N_R}$, then for an arbitrary $N_R \times N_R$ matrix
${\bf{F}}_k$, it can be written as
\begin{align}
\label{app_F}
{\bf{F}}_k&={\bf{M}}_{k}^{-\frac{1}{2}}{\bf{U}}_{{\boldsymbol{\Theta}}_k}{\boldsymbol
\Sigma}_{{\bf{F}}_k}{\bf{U}}_{{\bf{T}}_k}^{\rm{H}}
{\bf{R}}_{{\bf{x}}_k}^{-\frac{1}{2}}
\end{align} where
the inner matrix ${\boldsymbol \Sigma}_{{\bf{F}}_k}$ equals to
${\boldsymbol
\Sigma}_{{\bf{F}}_k}={\bf{U}}_{{\boldsymbol{\Theta}}_k}^{\rm{H}}{\bf{M}}_{k}^{\frac{1}{2}}{\bf{F}}_k{\bf{R}}_{{\bf{x}}_k}^{\frac{1}{2}}
{\bf{U}}_{{\bf{T}}_k}$.

Putting (\ref{app_F}) into (\ref{equ:L_1}), and with the following
definitions (the same as the definitions in (\ref{case_2}) and
(\ref{case_3}))
\begin{align}
{\bf{M}}_{k}^{-\frac{\rm{H}}{2}}{\bf{\hat
H}}_{rd,k}^{\rm{H}}{\bf{\hat H}}_{rd,k}{\bf{M}}_{k}^{-\frac{1}{2}}&=
{\bf{U}}_{{\boldsymbol{\Theta}}_k}{\boldsymbol
{\Lambda}}_{{\boldsymbol{\Theta}}_k}{\bf{U}}_{{\boldsymbol{\Theta}}_k}^{\rm{H}},
\\
 {\bf{R}}_{{\bf{x}},k}^{-\frac{1}{2}}{\bf{\hat
H}}_{sr,k}{\bf{R}}_{s,k} &={\bf{U}}_{{\bf{T}}_k}{\boldsymbol
\Lambda}_{{\bf{T}}_k}{\bf{V}}_{{\bf{T}}_k}^{\rm{H}},
\end{align}
the equalizer ${\bf{G}}_k$ can be reformulated as
\begin{align}
\label{app_G} &{\bf{G}}_k\nonumber \\
=& {\bf{R}}_{{\bf{s}}_k}({\bf{\hat
H}}_{rd,k}{\bf{F}}_k{\bf{\hat H}}_{sr,k})^{\rm{H}}({\bf{\hat
H}}_{rd,k}{\bf{F}}_k{\bf{R}}_{{\bf{x}}_k}{\bf{F}}_k^{\rm{H}}
{\bf{\hat
H}}_{rd,k}^{\rm{H}}+\eta_k{\bf{I}}_{M_D})^{-1} \nonumber \\
=&({\bf{R}}_{{\bf{x}}_k}^{-\frac{1}{2}}{\bf{\hat
H}}_{sr,k}{\bf{R}}_{{\bf{s}}_k})^{\rm{H}}(
{\bf{R}}_{{\bf{x}}_k}^{\frac{1}{2}}{\bf{F}}_k^{\rm{H}} {\bf{\hat
H}}_{rd,k}^{\rm{H}}{\bf{\hat
H}}_{rd,k}{\bf{F}}_k{\bf{R}}_{{\bf{x}}_k}^{\frac{1}{2}}+\eta_k{\bf{I}}_{M_R})^{-1}\nonumber \\& \times {\bf{R}}_{{\bf{x}}_k}^{\frac{1}{2}}{\bf{F}}_k^{\rm{H}}{\bf{\hat
H}}_{rd,k}^{\rm{H}} \nonumber \\
 =&{\bf{V}}_{{\bf{T}}_k}\underbrace{{\boldsymbol {
\Lambda}}_{{\bf{T}}_k}^{\rm{H}}( {\boldsymbol
\Sigma}_{{\bf{F}}_k}^{\rm{H}}{\boldsymbol {\Lambda}}_{{\boldsymbol
\Theta}_k}{\boldsymbol
\Sigma}_{{\bf{F}}_k}+\eta_k{\bf{I}}_{M_R})^{-1}
{\boldsymbol{\Sigma}}_{{\bf{F}}_k}^{\rm{H}}}_{\triangleq
{\boldsymbol \Sigma}_{{\bf{G}}_k}}
{\bf{U}}_{{\boldsymbol{\Theta}}_k}^{\rm{H}}{\bf{M}}_{k}^{-\frac{\rm{H}}{2}}{\bf{\hat
H}}_{rd,k}^{\rm{H}},
\end{align} where the second equality is due to the matrix
inversion lemma.

Putting (\ref{gamma}) from Appendix~\ref{Lemma_1} into
(\ref{equ:L_2}), after multiplying both sides of (\ref{equ:L_2})
with ${\bf{M}}_k^{-\frac{1}{2}}$, we have
\begin{align}
\label{106}
 & {\bf{M}}_k^{-\frac{1}{2}}{\bf{\hat
H}}_{rd,k}^{\rm{H}}{\bf{G}}_k^{\rm{H}}{\bf{G}}_k{\bf{\hat
H}}_{rd,k}{\bf{F}}_k{\bf{R}}_{{\bf{x}}_k}^{\frac{1}{2}}+
{\bf{M}}_k^{\frac{1}{2}}{\bf{F}}_k{\bf{R}}_{{\bf{x}}_k}^{\frac{1}{2}}\frac{\gamma_k}{\sigma_{n_2}^2}\nonumber \\
&={\bf{M}}_k^{-\frac{1}{2}}\left ({\bf{\hat
H}}_{sr,k}{\bf{R}}_{s,k}{\bf{G}}_k{\bf{\hat
H}}_{rd,k}\right)^{\rm{H}}{\bf{R}}_{{\bf{x}}_k}^{-\frac{1}{2}}.
\end{align} Then substituting ${\bf{F}}_k$ in (\ref{app_F}) and ${\bf{G}}_k$ in (\ref{app_G}) into (\ref{106}), we have
\begin{align}
\label{app_lambda}{\boldsymbol \Sigma}_{{\bf{F}}}=({\boldsymbol
\Lambda}_{{\boldsymbol \Theta}_k}{\boldsymbol
\Sigma}_{{\bf{G}}_k}^{\rm{H}}{\boldsymbol
\Sigma}_{{\bf{G}}_k}{\boldsymbol \Lambda}_{{\boldsymbol
\Theta}_k}+\frac{\gamma_k}{\sigma_{n_2}^2}{\bf{I}}_{N_R})^{-1}({\boldsymbol
\Lambda}_{{\bf{T}}_k}{\boldsymbol \Sigma}_{{\bf{G}}_k}{\boldsymbol
\Lambda}_{{\boldsymbol \Theta}_k})^{\rm{H}}.
\end{align}
Since ${\boldsymbol \Lambda}_{{\bf{T}}_k}$ and ${\boldsymbol
\Lambda}_{{\boldsymbol{\Theta}}_k}$ are rectangular diagonal
matrices (denoting their ranks by $p_k$ and $q_k$ respectively),
based on (\ref{app_lambda}), it can be concluded that ${\boldsymbol
\Sigma}_{{\bf{F}}_k}$ has the following form
\begin{align}
\label{diag_F} {\boldsymbol \Sigma}_{{\bf{F}}_k}=\left[
{\begin{array}{*{20}c}
   {\boldsymbol{A}}_{{\bf{F}}_k} & {\bf{0}} \\
   {\bf{0}} &  {\bf{0}}\\
\end{array}} \right]_{N_R \times M_R},
\end{align} where ${\boldsymbol{
A}}_{{\bf{F}}_k}$ is of dimension $ q_k \times p_k$ and to be
determined. Furthermore, putting (\ref{diag_F}) into the definition
of ${\boldsymbol \Sigma}_{{\bf{G}}_k}$ in (\ref{app_G}), we have
\begin{align}
\label{diag_G} {\boldsymbol \Sigma}_{{\bf{G}}_k}=\left[
{\begin{array}{*{20}c}
   {\boldsymbol{A}}_{{\bf{G}}_k} & {\bf{0}} \\
   {\bf{0}}_{} &  {\bf{0}}\\
\end{array}} \right]_{N_S \times M_D},
\end{align} where $
{\boldsymbol{ A}}_{{\bf{G}}_k}$ is of dimension $ p_k \times q_k$,
and to be determined. Substituting (\ref{diag_F}) and (\ref{diag_G}) into (\ref{app_F}) and (\ref{app_G}), it can be concluded that
\begin{align}
&{\bf{F}}_{k}=(P_{r,k} {\boldsymbol\Psi}
_{rd,k}+\sigma_{n_2}^2{\bf{I}}_{N_R})^{-\frac{1}{2}}{\bf{U}}_{{\boldsymbol{\Theta}}_k,q_k}{\boldsymbol
A}_{{\bf{F}}_{k}}{\bf{U}}_{{\bf{T}}_k,p_k}^{\rm{H}} {\bf{R}}_{{\bf{x}}_k}^{-\frac{1}{2}} \label{F_app}, \\
&{\bf{G}}_{k}={\bf{V}}_{{\bf{T}}_k,p_k}{\boldsymbol
A}_{{\bf{G}}_k}{\bf{U}}_{{\boldsymbol{\Theta}}_k,q_k}^{\rm{H}}(P_{r,k}
{\boldsymbol\Psi}
_{rd,k}+\sigma_{n_2}^2{\bf{I}}_{N_R})^{-\frac{\rm{H}}{2}} {\bf{\hat
H}}_{rd,k}^{\rm{H}} \label{G_app},
\end{align} where
\begin{align}
&{\boldsymbol{A}}_{{\bf{G}}_k}= {\boldsymbol {\bar
\Lambda}}_{{\bf{T}}_k}^{\rm{H}}( {\boldsymbol
A}_{{\bf{F}}_k}^{\rm{H}}{\boldsymbol {\bar \Lambda}}_{{\boldsymbol
\Theta}_k}{\boldsymbol
A}_{{\bf{F}}_k}+\eta_k{\bf{I}}_{p_k})^{-1}
{\boldsymbol{A}}_{{\bf{F}}_k}^{\rm{H}}, \label{F_G}
\end{align} and ${\boldsymbol{\bar \Lambda}}_{{\bf{T}}_k}$ is the $p_k
\times p_k$ principal submatrix of
${\boldsymbol{\Lambda}}_{{\bf{T}}_k}$.

\section{Proof of Property 3}
\label{Lemma_3}
Taking the trace of both sides of (\ref{F_3}) and (\ref{G_3}), and noticing that the resultant two equations are the same, it is obvious that
\begin{align}
\label{F_G_app}
{\rm{Tr}}({\boldsymbol {A}}_{{\bf{G}}_k}{\boldsymbol {\bar
\Lambda}}_{{\boldsymbol{\Theta}}_k}{\boldsymbol
{A}}_{{\bf{G}}_k}^{\rm{H}})=\frac{\gamma_k}{\eta_k\sigma_{n_2}^2}{\rm{Tr}}({\boldsymbol
{A}}_{{\bf{F}}_k}^{\rm{H}}{\boldsymbol {A}}_{{\bf{F}}_k}).
\end{align}
On the other hand, substituting (\ref{G_app}) into (\ref{gamma}) in Appendix~\ref{Lemma_1}, we have
\begin{align}
&{\rm{Tr}}({\boldsymbol {A}}_{{\bf{G}}_k}{\boldsymbol {\bar
\Lambda}}_{{\boldsymbol{\Theta}}_k}{\boldsymbol
{A}}_{{\bf{G}}_k}^{\rm{H}})=\frac{\gamma_k}{\sigma_{n_2}^2}P_{r,k}. \label{Equ_P_2}
\end{align} Comparing (\ref{F_G_app}) and (\ref{Equ_P_2}), it follows that
\begin{align}
&\frac{1}{\eta_k}{\rm{Tr}}({\boldsymbol{A}}_{{\bf{F}}_k}^{\rm{H}}{\boldsymbol{A}}_{{\bf{F}}_k})=P_{r,k}.
\label{power_const}
\end{align}

For the objective function in the optimization problem (\ref{MSE_Opt_1}), substituting (\ref{F_P_2}) and (\ref{G_P_2}) into the MSE
expression in (\ref{MSE}), the MSE on the $k^{\rm{th}}$ subcarrier can be written as
\begin{align}
\label{MSE_1}
&{\rm{MSE}}_k({\bf{F}}_k,{\bf{ G}}_k) \nonumber \\
=&{\rm{Tr}}({\boldsymbol{\bar
\Lambda}}_{{\bf{T}}_k}^2(\frac{1}{\eta_k}{\boldsymbol {A
}}_{{\bf{F}}_k}^{\rm{H}}{\boldsymbol{\bar
\Lambda}}_{{\boldsymbol{\Theta}}_k}{\boldsymbol {A
}}_{{\bf{F}}_k}+{\bf{I}}_{p_k})^{-1})\nonumber \\
& +
\underbrace{{\rm{Tr}}({\bf{R}}_{{\bf{s}}_k})-{\rm{Tr}}({\bf{R}}_{{\bf{s}}_k}{\bf{\hat
H}}_{sr,k}^{\rm{H}}{\bf{R}}_{{\bf{x}}_k}^{-1}{\bf{\hat
H}}_{sr,k}{\bf{R}}_{{\bf{s}}_k})}_{\triangleq c_k},
\end{align}where $c_k$ is a constant part independent of ${\bf{F}}_k$.
Therefore, based on (\ref{power_const}) and (\ref{MSE_1}), the optimization problem (\ref{MSE_Opt_1}) becomes as
\begin{align}
\label{MSE_Opt_SS}
& \min_{{\boldsymbol {A }}_{{\bf{F}}_k}} \ \ \ \sum_k
{\rm{Tr}}({\boldsymbol{\bar
\Lambda}}_{{\bf{T}}_k}^2(\frac{1}{\eta_k}{\boldsymbol {A
}}_{{\bf{F}}_k}^{\rm{H}}{\boldsymbol{\bar
\Lambda}}_{{\boldsymbol{\Theta}}_k}{\boldsymbol {A
}}_{{\bf{F}}_k}+{\bf{I}}_{p_k})^{-1})+c_k \nonumber \\
& {\rm{s.t.}} \ \ \ \ \
\frac{1}{\eta_k}{\rm{Tr}}({\boldsymbol{A}}_{{\bf{F}}_k}^{\rm{H}}{\boldsymbol{A}}_{{\bf{F}}_k})=P_{r,k}, \nonumber \\
& \ \ \ \ \ \ \ \  \sum_k P_{r,k}=P_r.
\end{align}

For any given $P_{r,k}$, then the optimization problem (\ref{MSE_Opt_SS}) can be decoupled into a collection of the following sub-optimization problems
\begin{align}
\label{Opt_Final}
& \min_{{\boldsymbol {A }}_{{\bf{F}}_k}} \ \ \
{\rm{Tr}}({\boldsymbol{\bar
\Lambda}}_{{\bf{T}}_k}^2(\frac{1}{\eta_k}{\boldsymbol {A
}}_{{\bf{F}}_k}^{\rm{H}}{\boldsymbol{\bar
\Lambda}}_{{\boldsymbol{\Theta}}_k}{\boldsymbol {A
}}_{{\bf{F}}_k}+{\bf{I}}_{p_k})^{-1}) \nonumber \\
& {\rm{s.t.}} \ \ \ \ \
\frac{1}{\eta_k}{\rm{Tr}}({\boldsymbol{A}}_{{\bf{F}}_k}^{\rm{H}}{\boldsymbol{A}}_{{\bf{F}}_k})=P_{r,k},
\end{align} where the constant part $c_k$ is neglected.
For any two $M \times M$ positive semi-definite Hermitian matrices
${\bf{A}}$ and ${\bf{B}}$, we have ${\rm{Tr}}({\bf{A}}{\bf{B}}) \ge \sum_{i}
\lambda_{i}({\bf{A}}) \lambda_{M-i+1}({\bf{B}})$, where $\lambda_i({\bf{Z}})$ denotes the $i^{\rm{th}}$
largest eigenvalue of the matrix ${\bf{Z}}$ \cite{Marshall79}. Together with the fact that
elements of the diagonal matrix ${\boldsymbol{\tilde
\Lambda}}_{{\bf{T}}_k}$ are in decreasing order, the objective function of (\ref{Opt_Final}) is minimized, when $({\boldsymbol {A
}}_{{\bf{F}}_k}^{\rm{H}}{\boldsymbol{\bar
\Lambda}}_{{\boldsymbol{\Theta}}_k}{\boldsymbol {A
}}_{{\bf{F}}_k}/{\eta_k}+{\bf{I}}_{N_k})$ is a diagonal matrix
with the diagonal elements in decreasing order. The objective function can be rewritten as
\begin{align}
&{\rm{Tr}}({\boldsymbol{\bar
\Lambda}}_{{\bf{T}}_k}^2(\frac{1}{\eta_k}{\boldsymbol {A
}}_{{\bf{F}}_k}^{\rm{H}}{\boldsymbol{\bar
\Lambda}}_{{\boldsymbol{\Theta}}_k}{\boldsymbol {A
}}_{{\bf{F}}_k}+{\bf{I}}_{N_k})^{-1})\nonumber \\
&={\bf{d}}^{\rm{T}}({\boldsymbol{\bar
\Lambda}}_{{\bf{T}}_k}^2)\underbrace{{\rm{d}}((\frac{1}{\eta_k}{\boldsymbol
{A }}_{{\bf{F}}_k}^{\rm{H}}{\boldsymbol{\bar
\Lambda}}_{{\boldsymbol{\Theta}}_k}{\boldsymbol {A
}}_{{\bf{F}}_k}+{\bf{I}}_{N_k})^{-1})}_{\triangleq{\bf{b}}}\triangleq
{\boldsymbol f}({\bf{b}}),
\end{align} where ${\bf{d}}({\bf{Z}})$ denotes the vector which consists of the main diagonal elements of the matrix ${\bf{Z}}$.

It follows that ${\boldsymbol f}({\bf{b}})$ is a Schur-concave
function of ${\bf{b}}$ \cite[\textsl{3.H.3}]{Marshall79}. Then,
based on \cite[\textsl{Theorem 1}]{Palomar03}, the optimal
${\boldsymbol {A}}_{{\bf{F}}_k}$ has the following structure
\begin{align}
& {\boldsymbol{A}}_{{\bf{F}}_k,{\rm{opt}}}=\left[
{\begin{array}{*{20}c}
  {\boldsymbol{
\Lambda}}_{{\bf{F}}_k,{\rm{opt}}} & {\bf{0}}_{N_k,p_k-N_k} \\
   {\bf{0}}_{q_k-N_k,N_k} &  {\bf{0}}_{q_k-N_k,p_k-N_k}\\
\end{array}} \right], \label{F_f}
\end{align} where ${\boldsymbol{
\Lambda}}_{{\bf{ F}}_k, {\rm{opt}}}$  is a $N_k \times N_k$ diagonal
matrix to be determined, and $N_k={{\min}}(p_k,q_k)$.

Putting
(\ref{F_f}) into the definition of ${\boldsymbol{A}}_{{\bf{G}}_k,{\rm{opt}}}$ in (\ref{F_G}), the structure of the optimal
${\boldsymbol{A}}_{{\bf{G}}_k,{\rm{opt}}}$ is given by
\begin{align}
{\boldsymbol{A}}_{{\bf{G}}_k,{\rm{opt}}}&=\left[
{\begin{array}{*{20}c}
  {\boldsymbol{
\Lambda}}_{{\bf{G}}_k,{\rm{opt}}} & {\bf{0}}_{N_k,q_k-N_k} \\
   {\bf{0}}_{p_k-N_k,N_k} &  {\bf{0}}_{p_k-N_k,q_k-N_k}\\
\end{array}} \right],
\label{G_g}
\end{align} where ${\boldsymbol{ \Lambda}}_{{\bf{G}}_k,{\rm{opt}}}$
is also a $N_k \times N_k$ diagonal matrix.

\end{document}